\documentclass[%
 a4paper, twocolumn,11pt
amsmath,amssymb
aps,floatfix,superscriptaddress,
]
{revtex4-2}
\pdfoutput = 1

\usepackage[utf8]{inputenc}
\usepackage[english]{babel}
\usepackage[T1]{fontenc}
\usepackage{xkeyval}
\usepackage{graphicx}
\usepackage[export]{adjustbox}
\usepackage{soul}
\usepackage{dcolumn}
\usepackage{bm}
\usepackage{epstopdf}
 \usepackage[usenames,dvipsnames]{pstricks}
 \usepackage{epsfig}
 \usepackage{pst-grad} 
 \usepackage{pst-plot} 
 \usepackage[space]{grffile} 
 \usepackage{etoolbox} 
 \makeatletter 
 \patchcmd\Gread@eps{\@inputcheck#1 }{\@inputcheck"#1"\relax}{}{}
 \makeatother
\usepackage{booktabs}
\usepackage{boxhandler}
\usepackage{amsmath}
\usepackage{subcaption}
\usepackage{braket}
\usepackage{amsmath}
\usepackage{float}
\makeatletter
\let\newfloat\newfloat@ltx
\makeatother
\usepackage{algorithm}
\usepackage{algpseudocode}

\usepackage{hyperref}
\hypersetup{
  colorlinks   = true,    
  urlcolor     = blue,    
  linkcolor    = blue,    
  citecolor    = blue      
}

\usepackage{multirow}
\usepackage{enumitem}

\begin{document}

\title{The closed-branch decoder for quantum LDPC codes}

\author{Antonio {deMarti iOlius}}
\email{ademartio@tecnun.es}
\affiliation{Department of Basic Sciences, Tecnun - University of Navarra, 20018 San Sebastian, Spain.}
\author{Josu {Etxezarreta Martinez}}
\email{jetxezarreta@tecnun.es}
\affiliation{Department of Basic Sciences, Tecnun - University of Navarra, 20018 San Sebastian, Spain.}


\begin{abstract}

Quantum error correction is the building block for constructing fault-tolerant quantum processors that can operate reliably even if its constituting elements are corrupted by decoherence. In this context, real-time decoding is a necessity for implementing arbitrary quantum computations on the logical level. In this work, we present a new decoder for Quantum Low Density Parity Check (QLDPC) codes, named the closed-branch decoder, with a worst-case complexity loosely upper bounded by $\mathcal{O}(n\text{max}_{\text{gr}}\text{max}_{\text{br}})$, where $\text{max}_{\text{gr}}$ and $\text{max}_{\text{br}}$ are tunable parameters that pose the accuracy versus speed trade-off of decoding algorithms. For the best precision, the $\text{max}_{\text{gr}}\text{max}_{\text{br}}$ product increases exponentially as $\propto dj^d$, where $d$ indicates the distance of the code and $j$ indicates the average row weight of its parity check matrix. Nevertheless, we numerically show that considering small values that are polynomials of the code distance are enough for good error correction performance. The decoder is described to great extent and compared with the Belief Propagation Ordered Statistics Decoder (BPOSD) operating over data qubit, phenomenological and circuit-level noise models for the class of Bivariate Bicycle (BB) codes. The results showcase a promising performance of the decoder, obtaining similar results with much lower complexity than BPOSD when considering the smallest distance codes, but experiencing some logical error probability degradation for the larger ones. Ultimately, the performance and complexity of the decoder depends on the product $\text{max}_{\text{gr}}\text{max}_{\text{br}}$, which can be considered taking into account benefiting one of the two aspects at the expense of the other.


\end{abstract}

\maketitle
\section{Introduction}
Quantum computing stands as one of the pillars of the next generation computing paradigm due to the theoretical promise of being able to run algorithms that can efficiently solve hard computational problems outside the reach of traditional computers \cite{algorithms}. Nevertheless, the unavoidable existence of decoherence introduces errors in the quantum computations that prevents us from accurately using quantum processors for obtaining the stupendous results promised by the theory. Specifically, qubits interact with their surrounding environment in ways we cannot prevent nor control, altering the quantum information being processed and, thus, resulting in utter failure \cite{tvqc}. Against this backdrop, the paradigm of quantum error correction codes (QECCs) has been proposed by the scientific community for dealing with such faulty qubits. Specifically, a QECC consists of $n$ of noisy physical qubits that are used to store the quantum information of a lower number, $k$, of logical or ``noiseless'' qubits \cite{gottesman}. Once the quantum information is encoded, partial information of the errors that may have occurred is extracted without destroying the quantum state, which results in a vector of classical information commonly known as syndrome. Then, the syndrome is fed to a classical algorithm named decoder, which is in charge of estimating which operation corrupted the encoded quantum information \cite{decoders}. If the recovered error corrects the actual error experienced by the qubits within the code, the original information is recovered successfully \cite{gottesman}. Therefore, decoding algorithms are an integral part of any quantum error correction code.

Several families of QECCs have been proposed since their conception by Shor in 1995 \cite{shorQEC}, namely, the most popular are the so-called surface codes due to their locality and large threshold probabilities \cite{kitaev,fowlerReview,decoders,xzzx}. While the probability threshold indicates the noise tolerance of the code when it operates over an error model, locality is specially relevant within the experimental field, since the physical qubits only interact with their nearest neighbours. Such property makes surface codes specially interesting for the mainstream superconducting or spin qubit platforms for which the qubits are spatially located over the chip and have restricted connectivity \cite{conncetSuper}. Importantly, the first experimental realizations of QECCs in real hardware have come in the form of surface codes \cite{wallraffSurf,googleSurf,LukinQEC}. Nevertheless, such architectural constraint comes at a cost of a vanishing coding rate and an excessive resource consumption making surface codes somehow inefficient. Quantum low density parity check codes (QLDPCs) represent one of the most interesting families of codes, specially after breakthrough results proving that good QLDPC codes in fact exist \cite{panteleev1,panteleev2}, and the recent proposal of Bivariate Bicycle (BB) codes that present quasi-local properties (with some long range interactions required per check) \cite{bravyidecoder}. Furthermore, recent development of quantum processors based on technologies that allow for full connectivity, e.g. neutral atoms, hint that the flourishing of experimental realizations of QLDPC codes might be nearer than expected \cite{LukinQEC,bonillaLDPC}.

Moreover, surface codes have an additionally interesting feature, they can be decoded through the consideration of a graph to which attempt to find a minimum weight perfect matching (MWPM) \cite{dennis, wuinterpretation}. That is, given a graph and a set of vertices which are considered as non-trivial, finding the minimum set of edges that matches all non-trivial vertices. This problem has been studied in depth within the QEC community and many advances have been made to solve it efficiently via the Blossom algorithm \cite{blossom1, dennis, fowlero1, fowler2} or the Union Find (UF) algorithm \cite{UF, peeling}. Furthermore, techniques involving the growth of clusters in order to solve the Blossom problem have been recently introduced, where non-trivial elements of the syndrome were considered to grow radially until encountering another growing non-trivial cluster, lowering the average complexity of the decoder to a great extend and contributing significantly to the overall goal of having real-time decoding in surface codes \cite{sparseblossom, fussionblossom}. Unfortunately, this significant advances for the decoding of the surface code cannot be applied to more general quantum error correcting codes, such as QLDPC codes, due to their non-matching structure. For QLDPCs, the decoder of choice is the Belief Propagation Ordered Statistics Decoder (BPOSD) \cite{panteleev1}. Albeit its good general performance, BPOSD suffers from a large complexity which may not make it a suitable real-time decoding method.  
\begin{figure}
    \centering
    \fbox{\includegraphics[width = .2\textwidth]{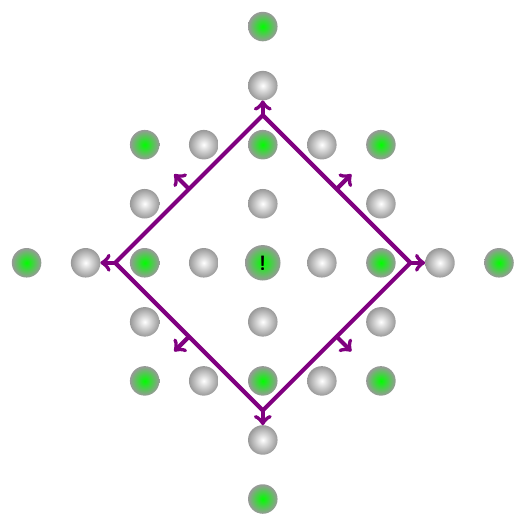}}
    \fbox{\includegraphics[width = .2\textwidth]{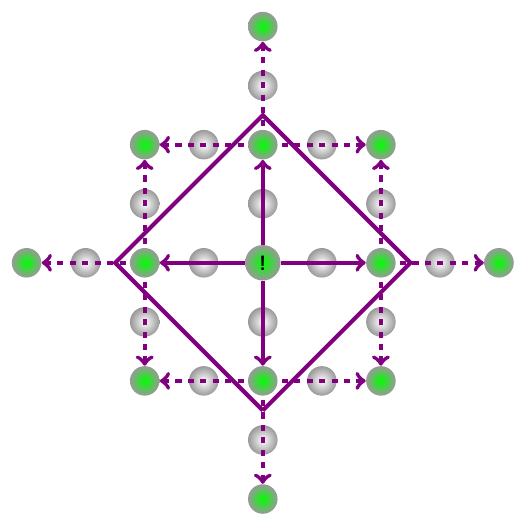}}
    \caption{Comparison between the cluster growth of a non-trivial syndrome element of a planar surface code considering a conventional matching decoder and the CB decoder. }
    \label{clustergrowths}
\end{figure}

In this work, we introduce a new decoding algorithm for QLDPC codes reminiscent to the matching decoding algorithms with a complexity dominated by two parameters that can be capped to an arbitrary extent by the user at the expense of its performance, the Closed-Branch decoder (CB decoder). The CB decoder considers cluster growth in a similar manner than the aforementioned UF \cite{UF} and Sparse Blossom \cite{sparseblossom} decoding methods. Nevertheless, the cluster does no longer consist of an overall circle where the initial non-trivial element is in the center. We consider the growth from the point of view of non-trivial Pauli chains in what we will define as branch instances. Figure \ref{clustergrowths} illustrates the growth of a non-trivial element cluster considering matching decoders \cite{UF, sparseblossom} to the left and the growth considered in the CB decoder to the right. Notice how, for the matching decoder, the cluster grows in all directions while the CB decoder would consist on considering a number of vectors that grows in an exponential manner as the cluster increases its radius. This exponential growth may imply that the CB decoder is not a worthy competitor for matching decoding, but, as discussed in the text, one can establish bounds in the maximum number of paths to be considered that lay far from considering all possible ones and still obtain good error correction performances at low complexity. 


The article begins by introducing the notions of closed branches and closed trees in order to follow with a thorough definition of the closed branch decoder and its variants. Afterwards, a series of numerical results of the performance of the CB decoder for several codes from the family of Bivariate Bicycle (BB) codes \cite{bravyidecoder} will be presented. For this article, we will consider three different noise-model scenarios, depolarizing data qubit noise, phenomenological noise and circuit-level noise and compare it to the performance of the BPOSD decoder. The CB decoder shows little performance degradation compared to BPOSD for data qubit and phenomenological noise models, considering the pseudothresholds ($p = P_L$), $\approx 6\%$ and $\approx 4\%$, respectively, when the complexity is kept low (as a function of code distances). The circuit-level noise model results more challenging, and results in a greater logical error rate degradation. We observe that for the small $[[72,12,6]]$ BB code, the decoder can operate similar to BPOSD (at $0.2\%$ pseudothreshold). However, the larger codes BB codes, $[[108,8,10]]$ and $[[144,12,12]]$, experience a degradation of approximately two orders of magnitude when being decoded with the proposed BP+CB decoder for such model. Though, the pseudothresholds for those codes lay at $0.2\%$ for the BP+CB decoder, which is not a significant degradation in comparison to the $0.4\%$ achieved by the BPOSD decoder. Notably, this is obtained by considering the same decoder complexity as for the phenomenological noise, implying that the proposed CB post-processing method is significantly faster than the traditional OSD protocol. Therefore, we consider the BP+CB decoder a promising candidate for fast decoding of QLDPC codes by paying the toll of not being as accurate as its BPOSD counterpart. Note that decoding speed is extremely important for implementing magic state injection based fault-tolerant algorithms \cite{decoders,terhal} and, thus, the potential of the proposed decoder. The article will finish with a conclusion showcasing the achievements and future work which may follow.

\section{Closed branches and closed trees}

The closed-branch decoder involves categorizing the total error affecting the code into subsets of error mechanisms referred to as closed branches. For a Quantum Error Correction Code (QECC) and given measured syndrome \footnote{Note that in the QEC jargon, the elements of the syndrome are also referred as checks. In this sense, a non-trivial check refers to a non-zero syndrome element.}, a \textit{closed branch} is defined as a set of individual error mechanisms where the adjacent checks meet the following two conditions.

\begin{itemize}
    \item The set of checks adjacent to and odd number of error mechanisms within the set must be non-trivial.
    \item The set of checks adjacent to an even number of error mechanisms within the set must be trivial.
\end{itemize}

\begin{figure}
    \centering
    \includegraphics[width = 0.45\textwidth]{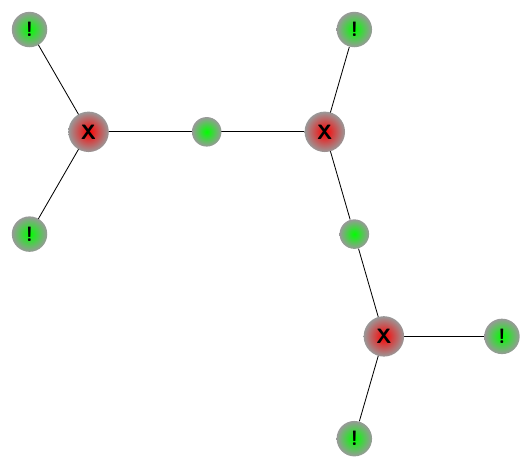}
    \caption{Closed branch encompassing three data qubits and eight checks. The red circles represent data qubits that have experienced bit-flip ($\mathrm{X}$) Pauli errors, green circles represent checks, while exclamation marks indicate non-triviality of a check. Black lines indicate adjacency or connectivity, i.e. non-triviality for the data-column check-row in the parity check matrix.}
    \label{3adj}
\end{figure}

Each individual error mechanism in a QECC triggers a set of checks. If a check is triggered by two distinct error mechanisms, it becomes trivial due to its binary nature. Therefore, a closed branch signifies an error instance that aligns with a portion of the syndrome. In Figure \ref{3adj}, a closed branch composed of three Pauli $\mathrm{X}$-errors is illustrated. Note how the non-trivial checks are only adjacent once to the branch while the trivial ones are adjacent to two data qubits satisfying the previous conditions. A closed branch suggests a potential local recovery based on a set of non-trivial elements within the syndrome, nevertheless, it does not guarantee a total recovery. For that to happen, one must take into account a second concept. Considering a QECC and a syndrome, a \textit{closed tree} is a set of closed branches which satisfy the conditions of a closed-branch while considering all non-trivial checks within the syndrome. A closed tree will always refer to an error which matches the syndrome. Nevertheless, such property does not imply that it is the most likely error to have occurred. Not all error patterns that correspond to the observed syndrome have the same effect on the code. Recovering an error that satisfies the syndrome conditions but belongs to a different logical error class will change the logical state of the code. In this sense, the decoder should attemt to solve the problem of finding the closed-tree that has the highest probability of occurrence. The existence of the so-called degeneracy makes the decoding problem in QEC to be different to the one in classical coding \cite{patDegen}. Therefore, and as a result of such difference, two decoding rules are usually studied in QEC: quantum maximum likelihood decoding (QMLD) and degenerate quantum maximum likelihood decoding (DQMLD) \cite{decoders, patDegen, hardness}. QMLD consists in determining the error pattern with higher probability of occurrence that matches the observed syndrome. This rule reduces to looking for the error with lowest weight that matches the syndrome whenever the noise model considers that the occurrence of the faults are independent among them. In this sense, QMLD is actually the classical decoding rule applied for QECCs. Differently, the DQMLD rule consists in attempting to find the most probable logical error class matching the measured syndrome. DQMLD is the optimal decoding rule for quantum stabilizer codes, but it is much more expensive in terms of computational complexity \footnote{Decoding stabilizer codes by taking into account all logical error classes is a \#P-complete problem \cite{hardness}.}, implying that in order to make practical decoders, the suboptimal QMLD rule is usually employed. We follow this logic for our decoding proposal and, thus, the decoding problem is reduced to the following goal: \textit{Find the lowest weight closed tree that matches a given error syndrome}.

\section{The closed branch decoder}

The CB decoder seeks to achieve the proposed goal of finding the lowest weight closed tree by considering branch instances and growing them until they become closed branches. We consider branch instances as any individual error mechanism adjacent to a number of non-trivial checks. As seen in the previous section, if the event is adjacent to any number of trivial checks, it does not satisfy the closed branch condition and, thus, additional events which also share the trivial check must be considered. We refer to the process of considering events adjacent to a specific trivial check from a branch as branch growth.

\subsection{Branch growth}

\begin{figure}
    \centering
    \fbox{\includegraphics[height = 0.3\textheight]{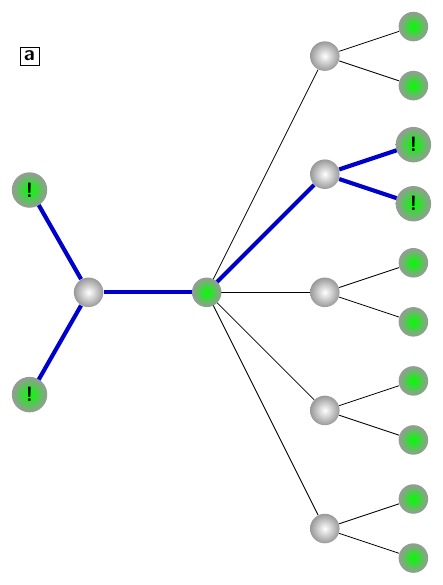}}
    \fbox{\includegraphics[height = 0.3\textheight]{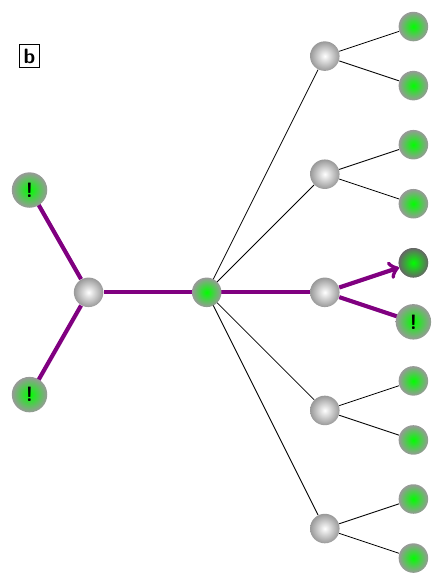}}
    \fbox{\includegraphics[height = 0.3\textheight]{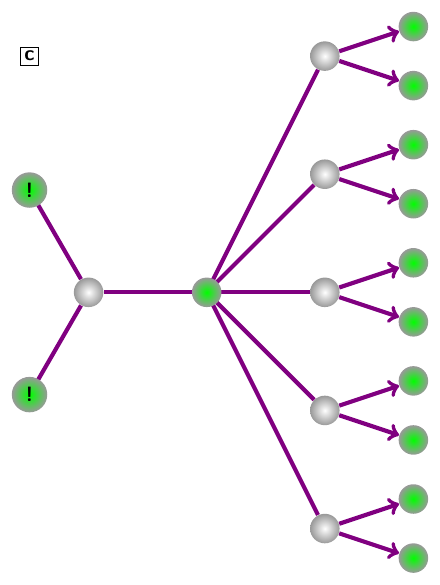}}
    \caption{Possible growths considering a branch instance composed by a data qubit adjacent to two non-trivial checks and a trivial check.}
    \label{growth}
\end{figure}

In order to grow a branch we need to consider the trivial checks that do not fulfill the closed branch conditions and then study their other adjacent individual error mechanisms. Figure \ref{growth} illustrates three different growth scenarios for a bivariate bicycle (BB) code \cite{bravyidecoder} branch instance composed by a data qubit error adjacent to two non-trivial checks under data qubit noise. The BB codes are a family of CSS codes where every data qubit is adjacent to three $\mathrm{X}$-checks and three $\mathrm{Z}$-checks and every $\mathrm{X}$ and $\mathrm{Z}$-check is adjacent to six data qubits. The initial branch instance considered is the left grey circle (data qubit), which is adjacent to two non-trivial checks and a trivial one. When considering a growth on that branch, we consider the other 5 data qubits adjacent to the trivial check as portrayed by the three illustrations. Each of those data qubits is itself adjacent to another two checks. If for any data qubit, its other checks are non-trivial, as illustrated in subfigure \ref{growth}\textbf{a}, then we recover a closed branch and the growth is completed. We call this phenomenon to \textit{close a branch}. On the other hand, if only one of the data qubits presents a single adjacency with a non-trivial check, then, upon following growths, the branch will direct itself towards the other trivial check adjacent to the aforementioned data qubit, as pointed by the violet arrow in subfigure \ref{growth}\textbf{b}. Ultimately, in the worst case scenario where all the adjacent checks to the adjacent data qubits are trivial, future growths should be done in all possible directions, as shown in subfigure \ref{growth}\textbf{c}. Then, in order to recover a closed branch, the branches emanating from two trivial checks adjacent to a same data qubit should be closed. Whenever this scenario happens, we say that there has been a \textit{separation} in the branch, since now closing the branch instance necessitates closing a larger number of branches. Whenever we consider a separation of that type, we will first continue to grow one of the two trivial checks from each of the data qubits. Once one of them is closed, we will have to return to the other check correspondent to the same data qubit and attempt to close it as well. For the illustrations in this article, a blue line will indicate a closed branch while a violet line will indicate a branch instance pointing towards the checks it will grow upon.

Separations are troublesome, since they imply a growing number of branches to close which can prove catastrophic for codes involving large connectivity or large error probabilities. Therefore, a growing schedule needs to be provided. Here, when growing a branch, only the growths implying the minimum number of separations will be considered. Going back to Figure \ref{growth}, given the scenario of a growth through a single check, if the branch can be closed through a second error mechanism, it will be closed. Otherwise, if it cannot be closed, it will continue to grow through the error mechanisms which contain the minimum number of trivial checks other than the one used to reach them. Moreover, if there is a separation and several adjacent error mechanisms have different numbers of adjacent checks, only the ones with minimum number of adjacent trivial checks will be considered for future growths.

\begin{figure}
    \centering
    \includegraphics[height = 0.3\textheight]{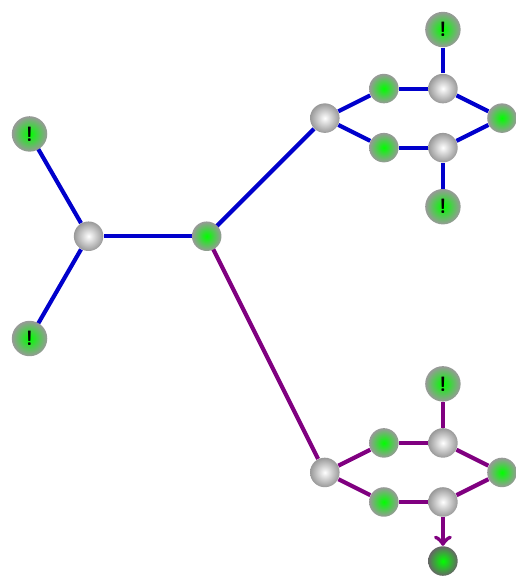}
    \caption{Loop instances while growing a branch.}
    \label{loop}
\end{figure}

Additionally, when growing branches, it must be considered that not all closed branches have a lineal structure such as the one from Figure \ref{3adj}. A separation in the growth of a branch can become a closed loop if one of the branches emanating from the separation grows towards the other trivial check. In Figure \ref{loop}, the previously considered growth results in the case of Figure \ref{growth}\textbf{c}, where all considered checks are trivial and, thus, a second growth is needed. The second growth results in the consideration of two loops. The top one closes the branch since both checks coincide in a check which is trivial while their other check is non-trivial, making the overall branch to satisfy the closed branch conditions. On the bottom, another loop occurs. This time, though, the closed branch property is not fulfilled due to the bottom check being trivial. Were there not the closed loop on top, the branch would continue to grow through the bottom trivial check, omitting the rest of them, since now the branch is no longer separated, i.e. there is only one trivial check to close.

\begin{figure}
    \centering
    \includegraphics[width = 0.35\textwidth]{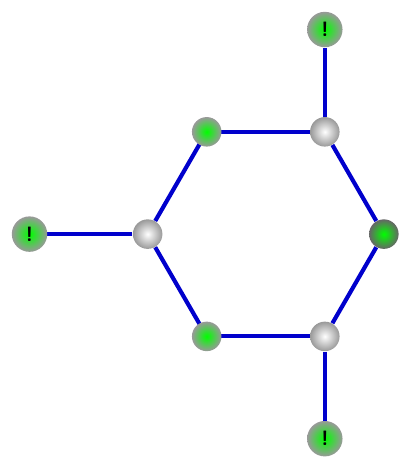}
    \caption{Closed branch where all error events within have two trivial checks.}
    \label{1detloop}
\end{figure}

Thus, through a set of ordered steps, a branch instance can be grown continuously until it reaches closure. 
After a number of growths or when we must consider too many branch instances due to many separations, if the branch has not closed it is discarded. Finding a good method for achieving relevant branch instances is of pivotal importance for the construction of the CB decoder. One may think that considering only branch instances where all checks are non-trivial but one may be a good convention to choose, since it avoids a separation on the first ever growth. Nevertheless, when considering loops, there can be closed branches where all their error mechanisms include more than one trivial checks, as in the example shown in Figure \ref{1detloop}. Thus, how many checks to search in the initial branch instance is a relevant parameter for the implementation of the CB decoder.

Moreover, the maximum number of branch growths and the maximum number of branches that we must consider are also parameters to take into account. A large number of growths to consider will allow the decoder to decode branches consisting of a large number of error mechanisms, while a large maximum number of branches will allow the decoder to consider a large number of separations. In this sense, the CB decoder is subjected to the usual accuracy-speed trade-off \cite{decoders}, but tuning in the parameters implies that the routine is flexible for improving the feature in which one is interested in. Importantly, as we will present in Section \ref{resultssec}, the CB decoder can operate with good accuracy for low values of those parameters, indicating that it can be very fast and present competitive error correcting capabilities.

\begin{figure}
    \centering
    \fbox{\includegraphics[width = 0.4\textwidth]{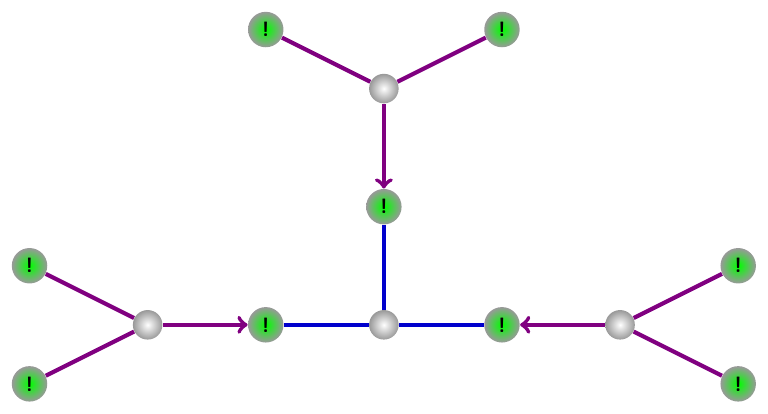}}
    \fbox{\includegraphics[width = 0.4\textwidth]{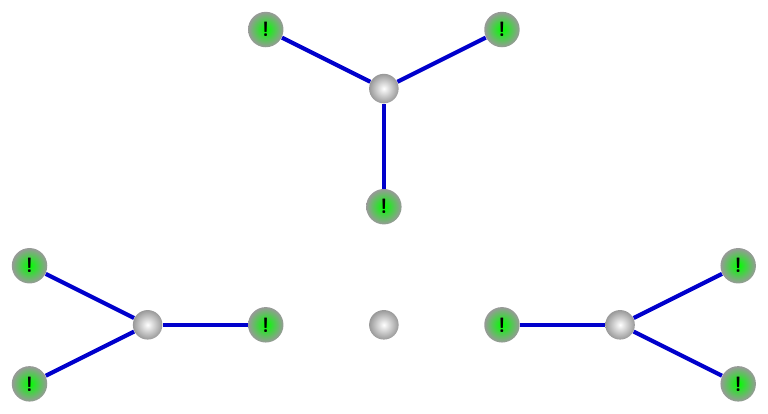}}
    \caption{The top image illustrates a scenario where a closed branch does not allow the creation of the closed tree of minimum growth under a single growth. In the second one a destructive growth of either of the periphery branch instances destroys the initial closed branch, allowing the construction of the closed tree.}
    \label{destruction}
\end{figure}

The CB decoder relies on growing branch instances until it closes them forming a closed tree of minimum weight, which represents the recovered error. Every time a growth of a branch instance results in a closed branch, the decoder considers all the non-trivial checks to be flipped to a trivial value and posterior branch growths will consider them as such. This may be troublesome, since this may make the overall necessary growths to increase or contribute to make the estimated recovery error to cause logical failure. In order to tackle this potential obstacle, the CB decoder considers two different types of growth: \textit{non-destructive} and \textit{destructive} growths. A non-destructive growth considers non-trivial checks belonging to closed branches as trivial. On the other hand, when a branch instance is growing destructively and encounters a non-trivial check belonging to a closed-branch which has been obtained non-destructively, it destroys such closed branch, making all its non-trivial checks to be considered as such once again. Figure \ref{destruction} portrays an instance where a closed branch consisted of a data qubit surrounded by non-trivial checks in the center prevents three branch instances from achieving the closed tree state. If we were restricted to only consider non-destructive growths, the first growth for any of the periphery branch instances would contribute to a separation. The second growth would close one of the outgoing branches from the center data qubit and the third growth would close the other one. Considering separations is expensive for the decoder. Thus, destructive growths are very convenient in this kind of situations. Under destructive growth, any of the periphery branch instances would destroy the central closed-branch, and the rest would close themselves equally, returning the closed tree in a single step. 

Up to this point and for the remainder of the article, we have described close branch decoding on the BB codes \cite{bravyidecoder}, where all data qubits have 3 adjacent $X$-checks and 3 adjacent $Z$-checks (A description of these codes is given on Appendix \ref{BBcodes}). Nevertheless, growing a branch on an arbitrary code under an arbitrary noise model can be done by constructing a parity check matrix where the columns represent the individual error mechanisms and the rows the syndrome elements, which we shall name noise parity check matrix. Notice that this definition is not necessarily the same as a the standard parity check matrix of a quantum code where columns represent $X$ and $Z$ operations on data qubits. Given a noise parity check matrix, every column would correspond to a potential closed branch instance. Given a branch instance column, the non-trivial adjacent checks will correspond to the non-trivial elements within the columns which correspond to non-trivial elements within the syndrome. Once a non-trivial element within the column correspondent to a trivial syndrome element is selected to grow, it will seek other non-trivial values horizontally, and consider other columns that also have a non-trivial value for that specific row. This will be repeated until the closed branch conditions are satisfied or the maximum number of branches allowed is reached. The closed branch conditions are generalized to a set of columns within the noise parity check matrix so that the set of rows to which the columns have a non-trivial value correspond to a syndrome value which is:

\begin{itemize}
    \item trivial if the row has a non-trivial value for an even number of columns within the closed branch.
    \item non-trivial if the row has a non-trivial value for an odd number of columns within the closed branch.
\end{itemize}

\subsection{The CB decoder schedule}

As defined in the previous subsection, the CB decoder will require a set of arguments which will constraint the exponential complexity growth of arbitrary branch instances. These will be the aforementioned maximum number of branches ($\text{max}_\text{br}$) to consider upon growing a branch instance, the maximum number of growths to consider ($\text{max}_\text{gr}$) and the number of maximum trivial checks that we can consider on an individual error mechanism that we may consider to grow ($\text{max}_\text{tcts}$). Given these arguments and a syndrome, the CB decoder acts guided by the following schedule:

It defines a value named "weight", which starts at a value of 2. Then, it iterates over the noise parity check matrix, and considers as closed branches all columns where all non-trivial rows correspond to non-trivial syndrome elements. Afterwards, it defines a value named "tcts" which starts at a 1 value, and corresponds to the numbers of trivial checks to search that an event must have in order to be considered as a branch instance. We then iterate over the noise parity check matrix columns again and, if we find a column which has a minimum of 1 non-trivial row and a number of "tcts" trivial ones, we consider growing it non-destructively "weight" times. For any growth, if the number of branches to consider increases to a value superior to $\text{max}_\text{br}$, the branch instance is rejected. After this is done, the process is repeated iteratively by increasing by one the value of "tcts" until "tcts" = $\text{max}_\text{tcts}$. Afterwards, if there are still non-trivial syndrome elements, the process is repeated but with destructive growths. If after the destructive growths the set tree has not been found, the value of "weight" is increased by 1 and the whole process is repeated. This is done until "weight"=$\text{max}_\text{gr}$ and, reached that point, if the problem is not resolved, the CB decoder fails.

\begin{figure}
    \centering
    \fbox{\includegraphics[width = 0.22\textwidth]{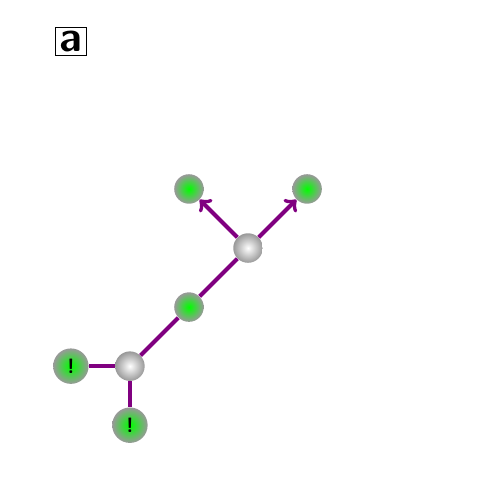}}
    \fbox{\includegraphics[width = 0.22\textwidth]{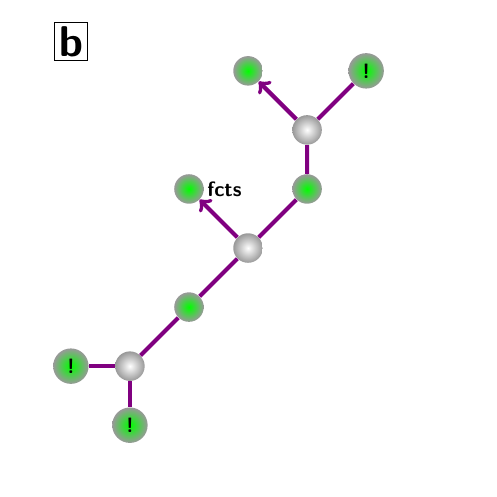}}
    \fbox{\includegraphics[width = 0.22\textwidth]{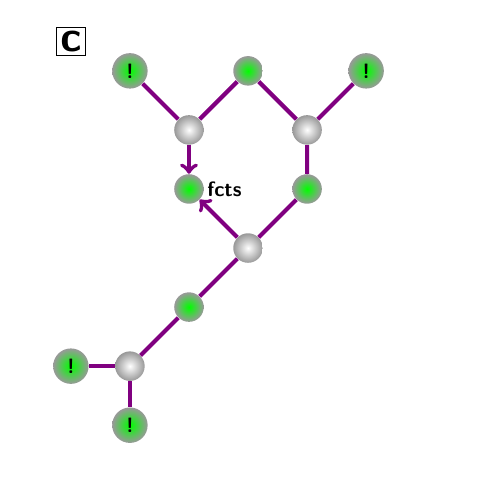}}
    \fbox{\includegraphics[width = 0.22\textwidth]{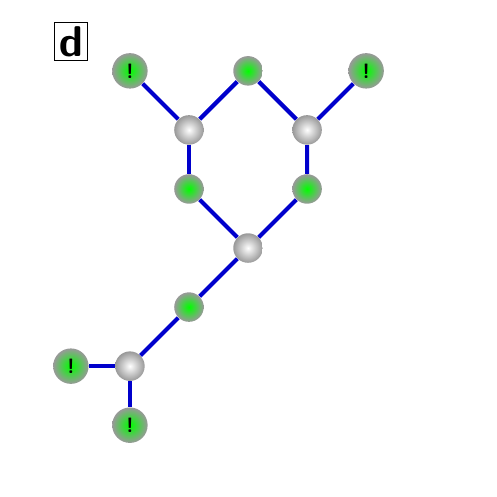}}
    \caption{The four images illustrate the closing of a branch instance through a loop detection event. In \textbf{a}, a separation is produced, where the trivial check on the bottom-left is adjacent to data qubits which themselves are only adjacent to trivial checks. In \textbf{b}, one of the aforementioned data qubits is considered by growing it through one of its adjacent checks while keeping the other as a future check to search (fcts). In \textbf{c}, a posterior growth reveals that the following check to search is the fcts, thus, we can close the branch as illustrated in \textbf{d}.}
    \label{loopcheck}
\end{figure}

In the unfortunate case in which a separation is produced, the branch will grow into all the adjacent error mechanisms. If all the error mechanisms have more than one trivial check to search, we will select one to grow while the remaining will be considered future checks to search (fcts). If a branch containing fcts is closed, it will consider an fcts to continue growing and will not be completely closed until there are no more fcts. Additionally, if while growing a branch grows upon a fcts, it will be closed producing a loop. As explained before, these loop events can be common in QLDPC codes and considering them can significantly reduce the complexity and accuracy of decoding using the ideas here. Thus, having a number of fcts is a necessary and beneficial condition for finding loops within the branch. Figure \ref{loopcheck} represents a scenario where a loop closes a branch instance which had a separation. Every time a branch instance with a number of fcts is grown, one must check if the next trivial check to search belongs to the fcts set.

Algorithm \ref{CBDecAlgo} summarizes the decoding process. Upon entering the for-loop, we define a weight, which sets the number of growths considered when considering the evolution of branch instances. Afterwards, a trivial cluster class is defined, a cluster class is a data object which stores closed branches obtained by destructive and non-destructive growths separately, non-trivial checks and error mechanisms involved in said closed-branches as attributes. Upon initialization, the cluster class is empty. We proceed searching for all columns within the noise parity check matrix, whichever closed-branches are detected will be stored within the cluster class with their correspondent non-trivial checks and error mechanisms. Afterwards, we will consider growing branch instances consisting of single error mechanisms with a number "tcts" of adjacent trivial checks. Then, the same is done but considering destructive growths. After every destructive growth considering a different number of trivial checks to search, we search weight-1 errors and grow branch instances non-destructively with one check to search to find any closed branch which may have been left as fallout from a larger closed-branch being destroyed. Afterwards, we verify if the non-trivial checks within the cluster correspond to all non-trivial checks within the syndrome and, were that to be the case, the error which corresponds to all the errors within the cluster is returned as the recovered error. If this condition is not achieved for all $\text{max}_{\text{gr}}$, the decoding process has failed and returns a trivial error.

\begin{algorithm*}
\caption{The CB Decoder}
\label{CBDecAlgo}
\begin{algorithmic}[1]
\Procedure{CB Decoding}{$\text{syndrome}$}
    \For{$\text{step} \gets 2$ \textbf{to} $\text{max}_{\text{gr}}$}
        \State $\text{weight} \gets \text{step}$
        \State $\text{cluster} \gets \text{Cluster\_class()}$
        \State $\text{cluster} \gets \text{weight\_1\_errors}(\text{syndrome}, \text{cluster})$
        \For{$\text{tcts} \gets 1$ \textbf{to} $\text{max}_{\text{tcts}}$}
            \State $\text{cluster} \gets \text{non\_dest\_branch\_growth}(\text{tcts},\text{cluster},\text{syndrome}, \text{weight},  \text{max}_{\text{br}})$
        \EndFor
        \For{$\text{tcts} \gets 1$ \textbf{to} $\text{max}_{\text{tcts}}$}
            \State $\text{cluster} \gets \text{dest}\_\text{branch}\_\text{growth}(\text{tcts},\text{cluster},\text{syndrome}, \text{weight},  \text{max}_{\text{br}})$
        \State $\text{cluster} \gets \text{weight\_1\_errors}(\text{syndrome}, \text{cluster})$
        \State $\text{cluster} \gets \text{non\_dest\_branch\_growth}(\text{tcts}_\text{arg} =1,\text{cluster}, \text{syndrome}, \text{weight},  \text{max}_{\text{br}})$
        \EndFor
    \If{$\text{cluster.checks} == \text{syndrome}$}
        \State \textbf{return} $\text{cluster.error}$
    \EndIf
    \EndFor
    \State \textbf{return} $\text{trivial\_eerror}$
\EndProcedure
\end{algorithmic}
\end{algorithm*}

\subsection{The BP+CB decoder}

The Belief Propagation decoder employs a message-passing algorithm suitable for addressing inference problems within probabilistic graphical models \cite{decoders}. For error correction codes, the BP algorithm consists in representing the parity check matrix of the code as a bipartite graph composed by two types of nodes, variable nodes and check nodes. Check nodes correspond to the rows in the parity check matrix while variable nodes correspond to the columns, edges correspond to non-trivial values within the matrix. Given a syndrome, and a noise model, the BP decoder returns a set of marginal probabilities for the check nodes which indicate the probability of them having undergone a physical error. Then, hard decisions can be made by making variable nodes with a marginal probability of being flipped above 1/2 are considered as physical errors. The BP decoder has shown excellent performance for decoding classical LDPC codes, but its performance for QLDPCs is far from optimal due to the presence of unavoidable $4$-loops (arising from the required commutation relationships required by stabilizer codes) and degeneracy \cite{panteleev1,patDegen}. This loss of performance is so significant that surface codes do not present a threshold when they are decoded by simply using BP \cite{decoders}. Nevertheless, due to its low complexity and the usefulness information the marginal probabilities provide, BP is considered as an interesting routine to aid other quantum decoders. Through the past years, it has been established that the belief propagation decoder can serve as a complexity free assistant to a number of QMLD decoders such as belief-matching \cite{beliefmatching, beliefmatchingcriger} or BPOSD \cite{panteleev1}. The CB decoder is no exception and its combination with BP results in a significant improvement on the average complexity in what we call the BP+CB decoder.

The BP+CB decoder begins by considering a BP decoding round on the syndrome through the noise parity check matrix. If the resulting marginal probabilities provide a recovered error which matches the syndrome, we return that error. Nevertheless, if the recovered error does not adjust to the syndrome, we run the CB decoder taking into account the computed marginal probabilities. This is done by reweighting CB decoder, i.e. by changing the weights of every error mechanism taking into account its \textit{log likelihood ratio} (llr), where $llr = \log(\frac{p_\mathrm{I}}{p_\mathrm{X}})$. Negative llrs imply that it is more likely that an error occurred and, thus, lower weight llrs using marginal probabilities indicate events likelier to belong to the closed branches which conform the overall error. Every error mechanism column $i$ is given a \textit{weight} provided by $\text{llr}_i -\text{min}(\text{llr}) + 1$, where $\text{min}(\text{llr})$ is the minimum llr value. This is done so as to make all weights positive while the minimum weight being equal to 1 will prevent infinite 0 weight loops. Now, instead of growing in all directions, the growth will be directed towards events with larger probability of occurrence i.e. lower llrs. This will be done by slightly changing the decoding mechanism, which can be seen in Algorithm \ref{CBPBDecAlgo}. This time, the growth parameter, "step", will begin to be 1 and establish a weight through $\text{weight} = \text{step}\times \text{max}(\text{llr} -\text{min}(\text{llr}) + 1)$. Henceforth, the procedure is the same as with the conventional CB decoder, except for the fact that branch instances are restricted to have a joint llr value below the value of "weight". Were this to happen the branch instance is discarded. Additionally, the step indicates the minimal number of growths which will be considered for every branch instance taking into account the weights from the error mechanisms. 

\begin{algorithm*}
\caption{The BP+CB Decoder}
\label{CBPBDecAlgo}
\begin{algorithmic}[1]
\Procedure{BP+CB Decoding}{$\text{syndrome}$}
    \State $\text{bp\_result} \gets \text{BP\_decoding}(\text{syndrome})$
    \If{$\text{bp\_result.syndrome} = \text{syndrome}$}
        \State \textbf{return} \text{bp\_result.error}
    \EndIf
    \State $\text{event\_weights} \gets \text{bp\_result.llrs} - \text{min}(\text{bp\_result.llrs}) + 1$
    \For{$\text{step} \gets 1$ \textbf{to} $\text{max}_{\text{gr}}$}
        \State $\text{weight} \gets \text{step} \times \text{max}(\text{event\_weights})$
        \State $\text{cluster} \gets \text{Cluster\_class()}$
        \State $\text{cluster} \gets \text{weight\_1\_errors}(\text{syndrome}, \text{cluster})$
        \For{$\text{tcts} \gets 1$ \textbf{to} $\text{max}_{\text{tcts}}$}
            \State $\text{cluster} \gets \text{non\_dest\_weighted\_branch\_growth}(\text{tcts},$ \State \hskip\algorithmicindent \hskip\algorithmicindent\hskip\algorithmicindent \hskip\algorithmicindent$\text{cluster}, \text{syndrome}, \text{weight}, \text{max}_{\text{br}}, \text{event\_weights})$
        \EndFor
        \For{$\text{tcts} \gets 1$ \textbf{to} $\text{max}_{\text{tcts}}$}
            \State $\text{cluster} \gets \text{dest\_weighted\_branch\_growth}(\text{tcts},$
            \State\hskip\algorithmicindent \hskip\algorithmicindent \hskip\algorithmicindent \hskip\algorithmicindent$\text{cluster}, \text{syndrome}, \text{weight}, \text{max}_{\text{br}}, \text{event\_weights})$
            \State $\text{cluster} \gets \text{weight\_1\_errors}(\text{syndrome}, \text{cluster})$
            \State $\text{cluster} \gets \text{non\_dest\_weighted\_branch\_growth}(\text{tcts\_arg}= $
            \State\hskip\algorithmicindent \hskip\algorithmicindent\hskip\algorithmicindent \hskip\algorithmicindent $1, \text{cluster}, \text{syndrome}, \text{weight},  \text{max}_{\text{br}}, \text{event\_weights})$
        \EndFor
        \If{$\text{cluster\_checks} = \text{syndrome}$}
            \State \textbf{return} $\text{cluster.error}$
        \EndIf
    \EndFor
    \State \textbf{return} $\text{trivial\_error}$
\EndProcedure
\end{algorithmic}
\end{algorithm*}

\subsection{Complexity, branch resolution and parallelization}

The complexity of the CB is dominated by the maximum number of branches and growths that we allow within the algorithm. Were the code to be unconstrained, for any branch instance in the worst case scenario, we would need to consider $k^{n_\text{growths}}$ branches for $n_\text{growths}$, where $k$ is the average number of non-trivial elements for the rows in the noise parity check matrix. Establishing a maximum number of branches limits this exponential growth on the number of branches. Ultimately, for a single branch instance considering that all of the $\text{max}_{\text{gr}}$ growths are limited to the generation of $\text{max}_{\text{br}}$ branches, the number of possible paths considered will be of the order of $\mathcal{O}(\text{max}_{\text{gr}}\text{max}_{\text{br}})$. As said before, this process will be done for all branch instances that will be a function of the code and the detected error mechanisms. The number of branch instances can be loosely upper bounded by the number of possible error mechanisms \footnote{Note that the number of possible branch instances cannot be higher than the number of error mechanisms. Considering that in order to start a branch instance some conditions must be met, then the actual worst case number of branch instances will be much smaller than $n$.}, $n$, resulting in the following upper bound for the complexity of the CB decoder $\mathcal{O}(n\text{max}_{\text{gr}}\text{max}_{\text{br}})$. We consider this upper bound not to be tight due to the loose upper bound of the maximum number of branch instances and the fact that loop search reduces the complexity significantly. Thus, the complexity of the decoder is capped by the parameters $\text{max}_{\text{gr}}$ and $\text{max}_{\text{br}}$. While we can lower these values arbitrarily, larger distances will require larger $\text{max}_{\text{gr}}$ values and denser noise parity check matrices (resulting from less sparse codes or from more complex error models) will require larger  $\text{max}_{\text{br}}$ values for obtaining good accuracy. Regarding the BP+CB decoder, the complexity will still be upper bounded by such expression since running a BP decoding round has complexity $\mathcal{O}(nj)$, where $j$ is the mean column-weight of the parity check matrix \cite{bpcompl}.

\begin{figure}[h]
    \centering
    \fbox{\includegraphics[width = 0.4\textwidth]{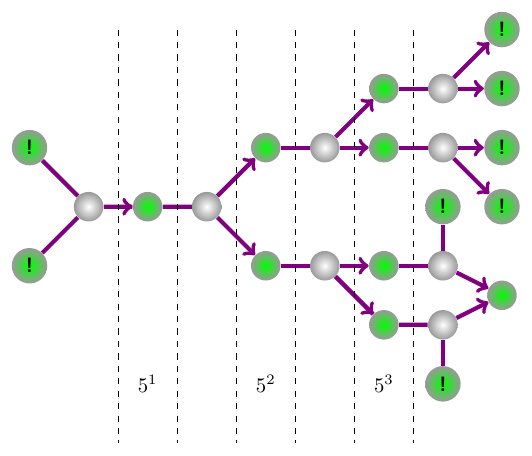}}
    \fbox{\includegraphics[width = 0.4\textwidth]{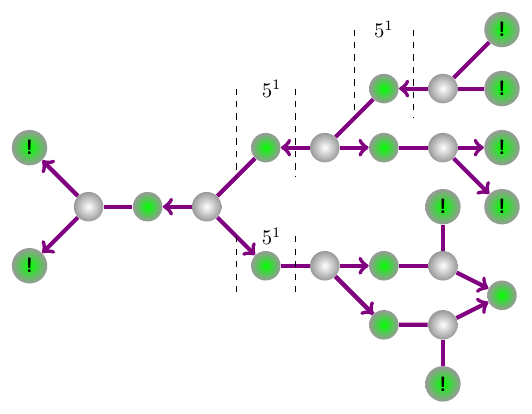}}
    \caption{Closed branch and two direction in which the decoder can solve it.}
    \label{brreal}
\end{figure}

For an arbitrary closed-branch to be successfully decoded, there are a two conditions which have to be satisfied. The decoder must be able to compute the closed-branch considering a number of branches below $\text{max}_{\text{br}}$ and a number of growths below $\text{max}_{\text{gr}}$ from at least one branch instance. Consider the closed branch in Figure \ref{brreal}, depending on the initial considered branch instance its decoding can require more or less complexity. In the top image, we consider growing the left branch instance, due to the fact that there are 3 growths in which no non-trivial checks are increased, the decoder must take into account $5^3 = 125$ branches until reaching the first closure for any check. On the other hand, if it grows from the top right branch instance as illustrated on the bottom of Figure \ref{brreal}, the number of branches to consider is quickly reduced after every separation considering that it might consider first the check which is adjacent to a closed event. This brings an interesting issue, where considering only a single branch outgoing from an error mechanism at a time may not be the most convenient choice \footnote{If the bottom image were to consider the left path after the first separation first and then the right one after the second, it would still require to process $5^3$ branches.}, but saves a lot of memory since considering all possible branch instances for all separations increases the number of branches, in the case of BB codes for data qubit noise, to $(5*2)^{n_{\text{sep}}}$.

Additionally, the iterations for different step values can be accelerated by considering their parallelization.  Recalling the process of the closed branch decoder, for each iteration, we consider a given branch weight at which branch instances are discarded. For the conventional decoder this is an integer value determining the weight of the branch while for the BP+CB one it depends on the overall weight of all error mechanisms within the branch instance. The first for loop in the pseudo code could be presented as a set of independent events running in parallel, once a number of closed trees would be retrieved, one could keep the one which had required the lowest amount of growths and present it as the recovered error.

\section{Results}\label{resultssec}
In this section, we numerically study the performance of the BP+CB decoder for bivariate bicycle codes (BB codes) \cite{bravyidecoder} over the three standard depolarizing noise models: pure data qubit noise, phenomenological noise and circuit-level noise. Within the BB codes, we will use three different codes with the properties indicated in TABLE.\ref{BBtable}. These specific codes are considered because of their similar properties due to their identical $m$ values and $A$ and $B$ polynomials. The difference in $l$ value results in different rates and distances. In Appendix \ref{BBcodes} we explain the process of generating BB code parity check matrices from the $l$, $m$, $A$ and $B$ parameters and in \ref{appNoise} we describe the three considered noise models in greater detail.

\begin{table}[htbp]
  \centering
  \caption{BB codes which will be considered for the simulations.}
  \label{BBtable}
  \begin{tabular}{|c|c|c|c|c|}
    \hline
    \textbf{$\bm{[[}n, k, d\bm{]]}$} & \textbf{$l$} & \textbf{$m$} & \textbf{$A$} & \textbf{$B$} \\
    \hline
    \hline
    $\bm{[[}72, 12, 6\bm{]]}$ & 6 & 6 & $x^3+y+y^2$ & $y^3 + x + x^2$ \\
    \hline
    $\bm{[[}108, 8, 10\bm{]]}$ & 9 & 6 & $x^3+y+y^2$ & $y^3 + x + x^2$  \\
    \hline
    $\bm{[[}144, 12, 12\bm{]]}$ & 12 & 6 & $x^3+y+y^2$ & $y^3 + x + x^2$  \\
    \hline
  \end{tabular}
\end{table}

The aim of this section is to numerically study the performance of the CB decoder. Nevertheless, these will be limited to the presented codes, noise models and $\text{max}_\text{gr}$, $\text{max}_\text{br}$ and $\text{max}_\text{tcts}$ values. For this study, $\text{max}_\text{gr}$, $\text{max}_\text{br}$ will be chosen in terms of the distance of the code and the number of non-trivial elements within the noise parity check matrix rows while $\text{max}_\text{tcts}$ will be $3$ for all simulations. Furthermore, the specific details of the numerical simulations are discussed in Appendix \ref{appA}.

\subsection{The closed branch decoder for pure data qubit noise}\label{datanoise}

\begin{figure}
    \centering
    \includegraphics[width = 0.5\textwidth]{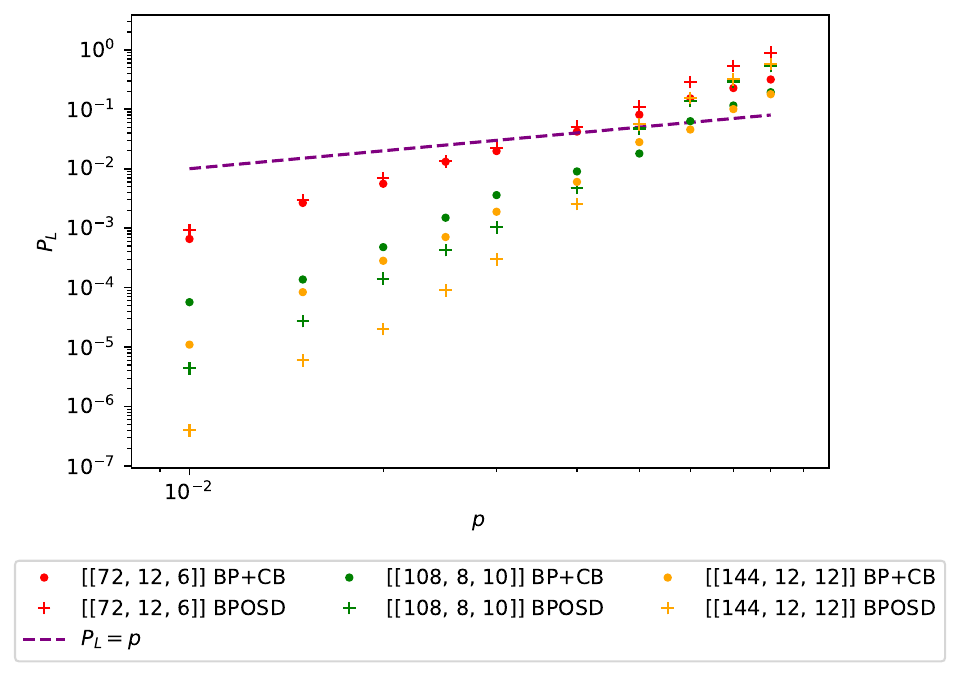}
    \caption{Logical error rate $P_L$ curves of different BB codes considering depolarizing data qubit noise under BP+CB and BPOSD-0 decoding with dependence on the physical error rate $p$. The dashed line illustrates the pseudo-threshold location through the physical probability range.}
    \label{deporesults}
\end{figure}

We begin by considering the depolarizing pure data qubit noise model. For this specific noise model, the noise parity check matrix is equal to the conventional parity check matrix of the code. Figure \ref{deporesults} illustrates the performance of the BP+CB decoder under depolarizing noise as opposed to BPOSD-0. For this specific case, we have considered $\text{max}_{\text{gr}} = 6$, $\text{max}_{\text{br}} = 10$ for all codes. The figure illustrates that the BPOSD and BP+CB curves are nearly identical for the smallest code, while the BPOSD yields a much better error correction performance, at low physical error rates, than the BP+CB for larger distance codes. Note that the pseudothresholds obtained are very similar for both methods, i.e. $\approx 6\%$. This feature can be explained from the point of view of the parameters at use. $\text{max}_{\text{gr}} = 6$, which is the minimum number of error mechanisms the decoder considers is equal to the distance of the smallest code, while equals only to the half of the distance for the largest code. Additionally, there are 6 non-trivial elements per row in the parity check matrix, and so the number of error mechanisms to consider per separation are 5. Thus, the number of emanating branches after $\text{max}_{\text{gr}}$ separations is $5^{\text{n}_{\text{gr}}}$. This exponential growth is greatly capped by our chosen $\text{max}_{\text{br}}$ favouring complexity as opposed to the decoding of errors consisting of closed branches with an elevated number of separations. This can be observed in the fact that codes with larger distances undergo worse BP+CB performance when compared to the BPOSD one. Larger distances imply the correctability of errors of larger weight, which may themselves be structured by a number of separations which would require a larger $\text{max}_{\text{br}}$ value, producing a failure in the BP+CB decoder. In this sense, we prove that the BP+CB decoder is able to present good decoding performance with a notably lower complexity than BPOSD, implying that it is much faster. Incrementing the branch parameters would make the decoder to be more precise at the cost of slowing down the process, indicating the flexibility of the proposed method.

The data qubit noise, although being the most well-known model, is also the most optimistic of the three which will be considered. Considering that the qubits acting as checks will not fail in the initiating, interacting or measuring processes required for syndrome extraction is overly optimistic considering the actual fault rates of those elements. For the BP+CB decoder, this consideration is very advantageous, since it involves considering a very sparse parity check matrix in which every row has 6 non-trivial columns, which consequently implies the aforementioned consideration of 5 error mechanisms upon every branch growth. This results in a good performance, even reaching same results for the smallest code, while retaining very low values of $\text{max}_{\text{gr}}$ and $\text{max}_{\text{br}}$ indicating a very low decoding complexity.

\subsection{The closed branch decoder for phenomenological noise}\label{phenomnoise}
We now consider a phenomenological depolarizing noise model (See Appendix \ref{appNoise}). The presence of a noisy syndrome requires a redefinition of the decoding process: instead of considering a single syndrome extraction which is used for recovering an error; $d$ syndrome extractions are done, where $d$ is the distance of the code. The overall syndrome extraction goes as follows: all the data qubits are initialized at the state $\ket{0}$, and a depolarizing channel in the data qubits is considered, the syndrome extractions proceeds ideally but for the measurement of the checks, where there is a probability of measurement failure, yielding a faulty syndrome. Afterwards, another data qubit depolarizing channel occurs followed by a noisy measurement syndrome extraction. This depolarizing on data, noisy measurement syndrome extraction process is repeated for $d$ rounds, finishing by measuring the data qubits in the $Z$-basis. The measurement of the data qubits serves as to consider a perfect measurement for $X$-checks. Measuring in the $Z$-basis implies the destruction of the effects of $Z$ error operators to the data qubits, thus, the decoder focuses on solving the $X$ errors which occurred through the $d$ decoding rounds while taking into account faulty measurements.

For phenomenological noise, the noise parity check matrix is the same one as the one used for data qubit noise with the addition of columns indicating the measurement error mechanisms. Every measurement error will produce two non-trivial checks in the overall syndrome and increases the number of non-trivial elements per row by 1 for the first and last measurement rounds and by two for the measurement rounds within the syndrome. 

\begin{figure}
    \centering
    \includegraphics[width = 0.5\textwidth]{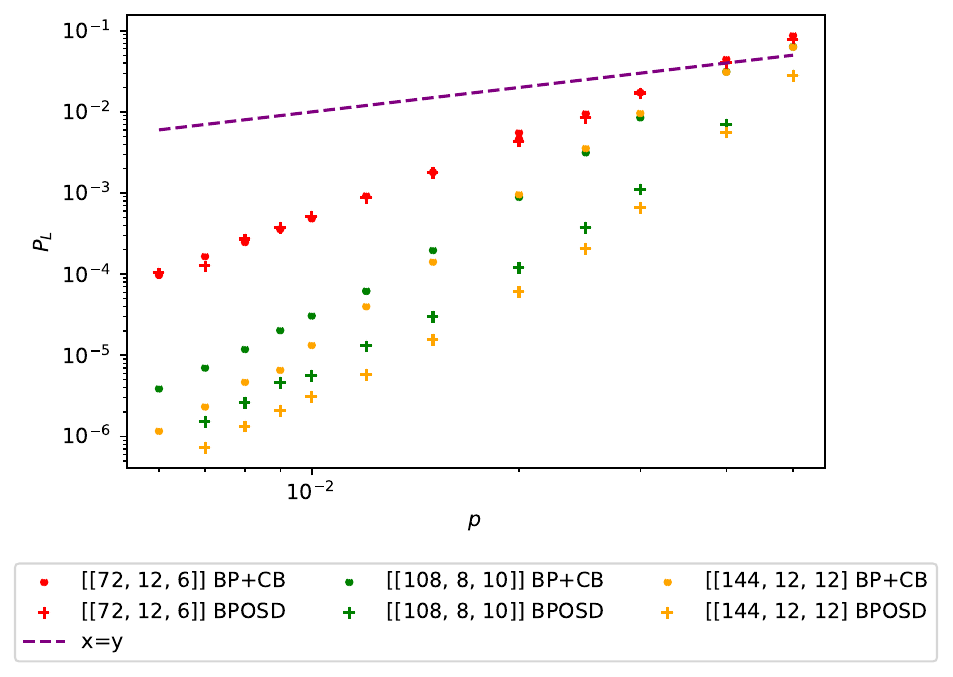}
    \caption{Logical error rate (per syndrome extraction cycle) $P_L$ curves of different BB codes considering phenomenological noise under BP+CB and BPOSD-0 decoding with dependence on the physical error rate $p$. The dashed line illustrates the pseudo-threshold location through the physical probability range.}
    \label{phenomresults}
\end{figure}

For this case we have considered  $\text{max}_{\text{gr}} = d$, $\text{max}_{\text{br}} = d^2$, where $d$ is the code distance. The results can be seen in Figure \ref{phenomresults}. Once again, the logical error curve of the BP+CB decoder matches the one of BPOSD for the smallest code, indicating that the established parameters suffice for good correction while being small when compared with the $\mathcal{O}(n^3)$ complexity of BPOSD. Note that the $n$ here refers to the size of the new noise parity check matrix and not to the code length. The other codes see a detriment in their performance when moving from BPOSD to CB+BP, although this time it is not as notable as in the pure data qubit depolarizing case, since we are considering larger $\text{max}_{\text{gr}}$ and $\text{max}_{\text{br}}$ for larger distance codes. Nevertheless, this does not suffice to reach a similar performance as the one obtained by the BPOSD decoder. Ultimately, the performance of the overall decoding process is somewhat compromised due to the fact that the parity check matrix to consider is larger and denser. Measurement errors increase the number of non-trivial columns for every syndrome element by one in the first and last measurement rounds and by two in the bulk measurement rounds, consequently increasing the number of error mechanisms per check to 8 and, as a result, increasing the number of possible branches to consider in the bulk to $7^{\text{n}_{\text{gr}}}$. As we are capping the number of maximum growths and branches to consider to $\text{max}_{\text{gr}}=d$ and $\text{max}_{\text{br}} = d^2$, the performance will be lowered. However, we consider that the speed boost obtained by the BP+CB decoder is very important while still presenting a reasonably good error correction performance.

\subsection{The closed-branch decoder for circuit-level noise}\label{clnnoise}
Circuit-level noise is the most realistic out of the three noise models which will be studied since it considers the errors for all faulty gates implicated in the syndrome extraction circuits: initialization, interaction and measurement. The standard circuit-level noise model establishes that all the elements of syndrome extraction circuits have probability $p$ of making the considered qubits interact with a non-trivial Pauli operator (See Appendix \ref{appNoise} for details). This implies an interesting disadvantage, since non-trivial Pauli operators can propagate through the CNOT gates in the syndrome extraction circuit. The consequence is unfortunate: single error mechanisms can produce errors of as much as weight $5$ when considering the phenomenological noise parity check matrix instead of the circuit-level noise one. The resulting noise parity check matrix can be obtained by means of the Stim \cite{stim} and beliefmatching \cite{beliefmatching} software packages. The result is a much larger noise parity check matrix, with $17$ non-trivial elements in the rows for the first and last syndrome extraction rounds and $34$ in the syndrome extractions within the bulk.

\begin{figure}
    \centering
    \includegraphics[width = 0.5\textwidth]{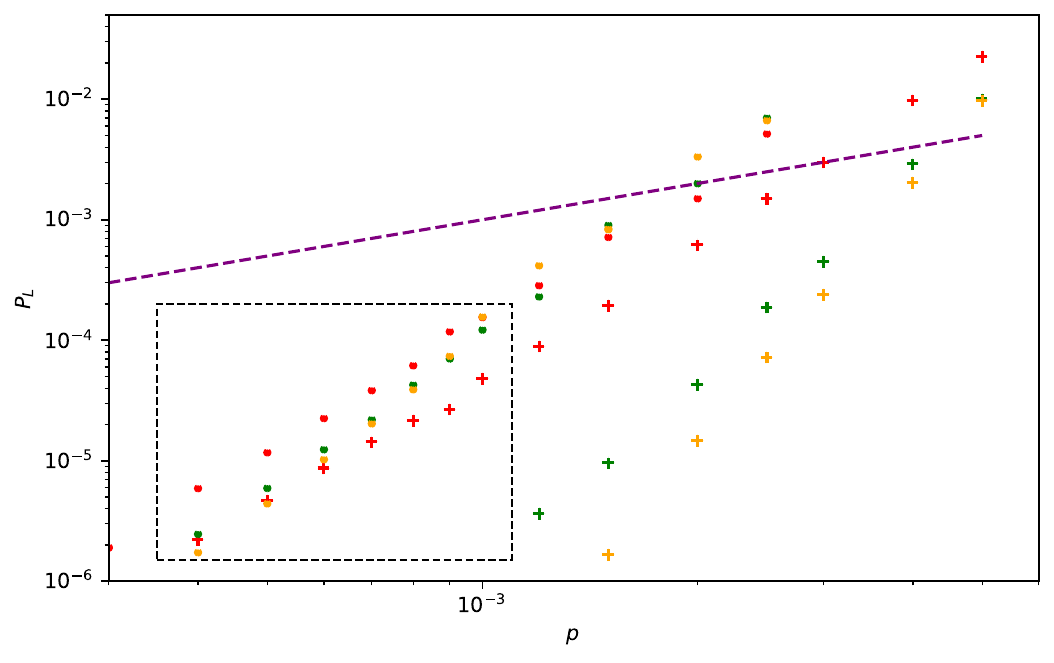}
    \includegraphics[width = 0.5\textwidth]{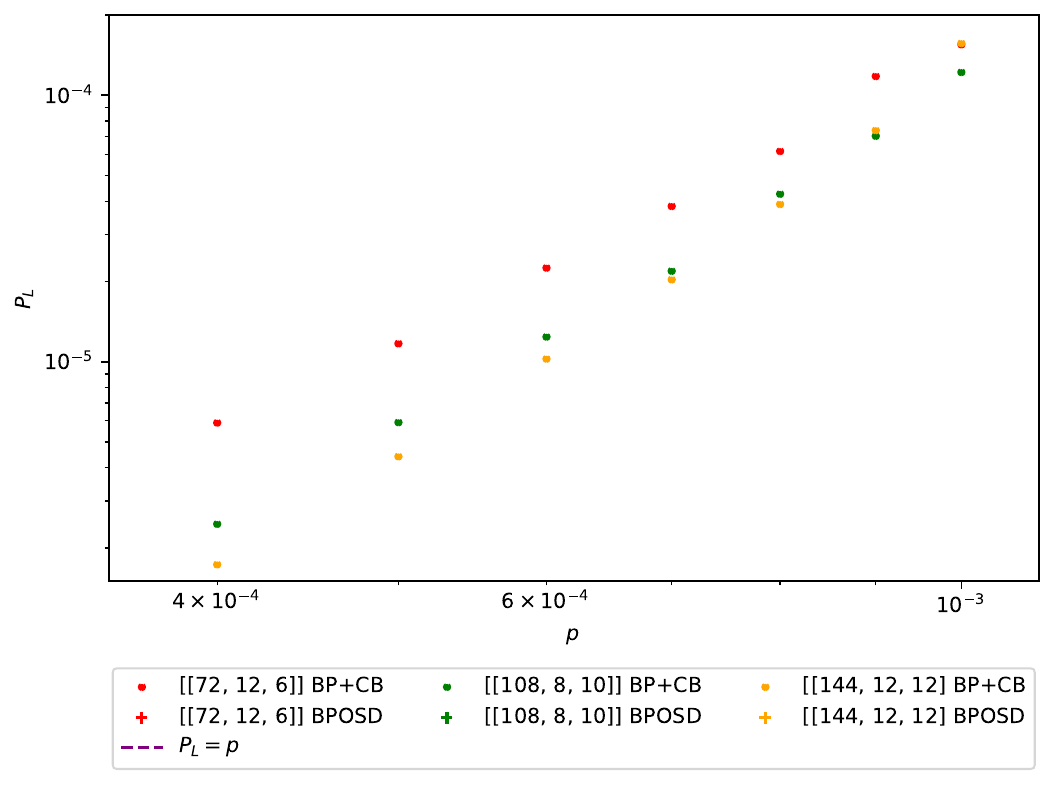}
    \caption{\textbf{(Top)} Logical error rate (per syndrome extraction cycle) $P_L$ curves of different BB codes considering circuit-level noise under BP+CB and BPOSD-0 decoding with dependence on the physical error rate $p$. The dashed line illustrates the pseudo-threshold location through the physical probability range. \textbf{(Bottom)} Magnified region pointed by dashed rectangle in top figure. It is intended for better visualization and thus the $[[72,12,6]]$ BPOSD result has been taken out. }
    \label{clnres}
\end{figure}

In order to study this noise model, at this point we consider the same BP+CB decoder instance as we did for the phenomenological noise, i.e. considering $\text{max}_{\text{gr}} = d$, $\text{max}_{\text{br}} = d^2$ parameters. The results can be seen in Figure \ref{clnres} where generally can be seen that the BP+CB decoder struggles to catch up with the error correction performance of BPOSD. Interestingly, the decoder is able to present little degradation when the smallest $[[72,12,6]]$ code is considered, even presenting a similar pseudothreshold around $\approx 0.2\%$. Regarding the largest codes, we can observe that the pseudothreshold is not significantly degraded when compared to BPOSD, i.e. it goes from $\approx 0.4\%$ to $\approx 0.2\%$. However, it can be seen that the performance worsens almost two orders of magnitude for those larger codes. Though, it can be seen that the BP+CB decoder is actually working under circuit-level noise for all three codes, improving the logical error rate as the code distance increases. The $[[108,8,10]]$ and $[[144,12,12]]$ codes perform in a pretty similar manner, achieving $P_L$ separation when lower physical error probabilities are considered. This separation can be seen at the lowest physical error probabilities for the BP+CB decoder (see Bottom figure in Figure \ref{clnres}), and should get more significant going lower, as for BPOSD. We do not further compute lower points due to the significant amount of computations required for those (See Appendix \ref{appA}). Notably, the most important thing to highlight is that this error rate is achieved at a significantly lower complexity than the one posed by BPOSD.


\section{Conclusions}
This article presents a novel quantum error correcting decoder incorporating concepts from surface code matching decoders to enable the decoding of a broader range of quantum error correction codes. The concepts of closed branches and closed trees have been introduced along with the closed-branch decoder and its variant, the belief propagation closed-branch decoder, in which belief propagation is introduced to improve the overall decoding process. In this sense, similar to BPOSD, the CB subroutine is executed only when BP fails to recover errors that match the syndrome. 

The main takeaway of the CB decoder consists in its competitive error correction capabilities at a much lower complexity than for BPOSD. Specifically, the complexity is dominated by $\mathcal{O}(n\text{max}_{\text{gr}}\text{max}_{\text{br}})$, where $\text{max}_{\text{gr}}$ and $\text{max}_{\text{br}}$ are tunable parameters. The selection of such parameters explicitly shows the accuracy versus speed trade-off usually present in decoding QECCs \cite{decoders}. In this sense, one can change the parameters to be faster or more precise by limiting its counterpart at the same time. Moreover, another necessary parameter to set is $\text{max}_{\text{tcts}}$, which corresponds to the  the initial separations that will be considered for initial branch instances. Moreover, the process can be parallelized by considering branch growths under different maximum weights independently. The lowest number of growths that achieves a closed tree is defined as the most probable recovered error.

The numerical results have shown the performance of the BP+CB decoder and BPOSD for the three standard depolarizing noise models. The BP+CB decoder has shown to present almost identical performance as BPOSD for the $[[72,12,6]]$ code for all noise levels, achieving very similar pseudothresholds with a lower complexity. For the other two larger BB codes, the BPOSD decoder has resulted to be more accurate, but the BP+CB decoder produced logical error curves which achieved the ones of BPOSD in the vicinity of the probability pseudothreshold for data qubit and phenomenological noise models requiring lower complexity. For circuit-level noise, the proposed BP+CB decoder is not sufficiently good to match the error correction performance of BPOSD for the larger codes. Specifically, the logical error rate of the largest BB codes considered seems to be degraded two orders of magnitude approximately. However, the pseudothresholds are in the vicinity of the ones achieved by the BPOSD decoder. Moreover, the BP+CB decoder obtains those results at a much lower complexity. This indicates that this decoder is potentially a good candidate for low latency decoding of QLDPC codes. This feature is specially relevant for the real-time decoding requirement of QECCs if fault-tolerant algorithms are targeted via magic state injection \cite{decoders,terhal}.

The results from this work open the possibility of research of a new decoder for QLDPC codes. While the structure of the decoder has been discussed, its performance for several classes of QECCs should be interesting follow up work. The obtained results indicate that its performance increases for sparse noise parity check matrices with low connectivity, indicating its suitability for other QLDPC codes than the BB codes considered in this article. Perhaps future research could follow these results into other noise models, such as the independent non-identically distributed noise model \cite{ton1, ton2, roffeinid}. Ultimately, it is of capital interest that QECCs to decode have low number of non-trivial values within the checks, which will imply a low number of emanating branches after a single separation. Studying methods for sparsifying dense noise parity check matrices so as to diminish their detriments on the BP+CB decoder could be an approach to tackle the results from circuit-level noise, as was done in \cite{beliefmatching}. The number of emanating branches after a separation will directly contribute to the necessary $\text{max}_{\text{br}}$ value for a satisfactory decoding process. While $\text{max}_{\text{gr}}\text{max}_{\text{br}} << n^2$, the process will be faster than the worst case complexity of BPOSD. A more elaborated study of the implications of $\text{max}_{\text{br}}$ and $\text{max}_{\text{gr}}$ with the performance is left as future work.

Furthermore, we consider that the CB decoder can further be parallelized in order to deal with all possible branch instances in parallel. Note that by doing this, the set of closed branches that is found may contain elements with overlapping non-trivial checks as the branch instances are grown independently. Therefore, this further parallelization would require to make a consensus among those possibly overlapping closed branches. If the final step is done with low complexity, this parallelization could further reduce the overall complexity of the BP+CB decoder to that of growing a single branch instance, i.e. $\mathcal{O}(\text{max}_{\text{gr}}\text{max}_{\text{br}})$. This could potentially make the BP+CB algorithm to be a decoder of even lower latency. We consider that the way in which the consensus should done is not trivial and we deem it as future work. 

Moreover, windowing techniques have been recently proposed in order to deal with the exponential backlog problem of decoding whenever matching decoders (MWPM and UF) are considered for the surface code \cite{browneSliding,TanSliding}. Considering that the backlog problem is a limiting factor for making arbitrary fault-tolerant computations (real-time decoding is required for magic state injection) \cite{decoders,terhal}, it is critical to tame it for other families of codes that have the potential to be integrated in quantum computers. The CB decoder has a ``matching''-like nature, implying that those windowing techniques can in principle be combined with the proposed decoder to deal with the exponential backlog problem for the more general code family of QLDPC codes. We also deem these studies as future work.

\section{Code availability}
At the current time, the authors are working on an open GitHub repository with the decoder implementation and its dependencies. We will reference the package in future versions of this work. At this point though, the raw code that supports the findings of this study is available upon reasonable request.

\section{Acknowledgements}
We warmly thank Dan Browne for hosting AdMiO in his group at UCL as well as for fruitful discussions and guidance of this research project. Moreover, we want to also acknowledge Oscar Higgott for the elaboration of a package for generating stim circuits specific to particular BB codes as well as for many useful comments and recommendations. Moreover, we would like to acknowledge Geoerge Umbrarescu for helping in the development of the decoder repository. We also thank Pedro Crespo for his guidance and the other members of the Quantum Information Group at Tecnun for their support.

This work was supported by the Spanish Ministry of Economy and Competitiveness through the MADDIE project (Grant No. PID2022-137099NBC44), by the Spanish Ministry of Science and Innovation through the proyect Few-qubit quantum hardware, algorithms and codes, on photonic and solidstate systems (PLEC2021-008251), and by the Ministry of Economic Affairs and Digital Transformation of the Spanish Government through the QUANTUM ENIA project call - QUANTUM SPAIN project, and by the European Union through the Recovery, Transformation and Resilience Plan - NextGenerationEU within the framework of the Digital Spain 2026 Agenda.

\bibliography{bibliography}

\begin{thebibliography}{47}%
\makeatletter
\providecommand \@ifxundefined [1]{%
 \@ifx{#1\undefined}
}%
\providecommand \@ifnum [1]{%
 \ifnum #1\expandafter \@firstoftwo
 \else \expandafter \@secondoftwo
 \fi
}%
\providecommand \@ifx [1]{%
 \ifx #1\expandafter \@firstoftwo
 \else \expandafter \@secondoftwo
 \fi
}%
\providecommand \natexlab [1]{#1}%
\providecommand \enquote  [1]{``#1''}%
\providecommand \bibnamefont  [1]{#1}%
\providecommand \bibfnamefont [1]{#1}%
\providecommand \citenamefont [1]{#1}%
\providecommand \href@noop [0]{\@secondoftwo}%
\providecommand \href [0]{\begingroup \@sanitize@url \@href}%
\providecommand \@href[1]{\@@startlink{#1}\@@href}%
\providecommand \@@href[1]{\endgroup#1\@@endlink}%
\providecommand \@sanitize@url [0]{\catcode `\\12\catcode `\$12\catcode `\&12\catcode `\#12\catcode `\^12\catcode `\_12\catcode `\%12\relax}%
\providecommand \@@startlink[1]{}%
\providecommand \@@endlink[0]{}%
\providecommand \url  [0]{\begingroup\@sanitize@url \@url }%
\providecommand \@url [1]{\endgroup\@href {#1}{\urlprefix }}%
\providecommand \urlprefix  [0]{URL }%
\providecommand \Eprint [0]{\href }%
\providecommand \doibase [0]{https://doi.org/}%
\providecommand \selectlanguage [0]{\@gobble}%
\providecommand \bibinfo  [0]{\@secondoftwo}%
\providecommand \bibfield  [0]{\@secondoftwo}%
\providecommand \translation [1]{[#1]}%
\providecommand \BibitemOpen [0]{}%
\providecommand \bibitemStop [0]{}%
\providecommand \bibitemNoStop [0]{.\EOS\space}%
\providecommand \EOS [0]{\spacefactor3000\relax}%
\providecommand \BibitemShut  [1]{\csname bibitem#1\endcsname}%
\let\auto@bib@innerbib\@empty
\bibitem [{\citenamefont {Montanaro}(2016)}]{algorithms}%
  \BibitemOpen
  \bibfield  {author} {\bibinfo {author} {\bibfnamefont {A.}~\bibnamefont {Montanaro}},\ }\bibfield  {title} {\bibinfo {title} {Quantum algorithms: an overview},\ }\href {https://doi.org/10.1038/npjqi.2015.23} {\bibfield  {journal} {\bibinfo  {journal} {npj Quantum Information}\ }\textbf {\bibinfo {volume} {2}},\ \bibinfo {pages} {15023} (\bibinfo {year} {2016})}\BibitemShut {NoStop}%
\bibitem [{\citenamefont {{Etxezarreta Martinez}}\ \emph {et~al.}(2021)\citenamefont {{Etxezarreta Martinez}}, \citenamefont {Fuentes}, \citenamefont {Crespo},\ and\ \citenamefont {Garcia-Frias}}]{tvqc}%
  \BibitemOpen
  \bibfield  {author} {\bibinfo {author} {\bibfnamefont {J.}~\bibnamefont {{Etxezarreta Martinez}}}, \bibinfo {author} {\bibfnamefont {P.}~\bibnamefont {Fuentes}}, \bibinfo {author} {\bibfnamefont {P.}~\bibnamefont {Crespo}},\ and\ \bibinfo {author} {\bibfnamefont {J.}~\bibnamefont {Garcia-Frias}},\ }\bibfield  {title} {\bibinfo {title} {Time-varying quantum channel models for superconducting qubits},\ }\href {https://doi.org/10.1038/s41534-021-00448-5} {\bibfield  {journal} {\bibinfo  {journal} {npj Quantum Information}\ }\textbf {\bibinfo {volume} {7}},\ \bibinfo {pages} {115} (\bibinfo {year} {2021})}\BibitemShut {NoStop}%
\bibitem [{\citenamefont {Gottesman}(1997)}]{gottesman}%
  \BibitemOpen
  \bibfield  {author} {\bibinfo {author} {\bibfnamefont {D.}~\bibnamefont {Gottesman}},\ }\emph {\bibinfo {title} {Stabilizer codes and quantum error correction}},\ \href@noop {} {\bibinfo {type} {Phd thesis}},\ \bibinfo  {school} {California Institute of Technology}, \bibinfo {address} {Pasadena, CA} (\bibinfo {year} {1997}),\ \bibinfo {note} {available at \url{https://thesis.library.caltech.edu/2900/2/THESIS.pdf}}\BibitemShut {NoStop}%
\bibitem [{\citenamefont {{deMarti iOlius}}\ \emph {et~al.}(2023{\natexlab{a}})\citenamefont {{deMarti iOlius}}, \citenamefont {Fuentes}, \citenamefont {Or\'us}, \citenamefont {Crespo},\ and\ \citenamefont {{Etxezarreta Martinez}}}]{decoders}%
  \BibitemOpen
  \bibfield  {author} {\bibinfo {author} {\bibfnamefont {A.}~\bibnamefont {{deMarti iOlius}}}, \bibinfo {author} {\bibfnamefont {P.}~\bibnamefont {Fuentes}}, \bibinfo {author} {\bibfnamefont {R.}~\bibnamefont {Or\'us}}, \bibinfo {author} {\bibfnamefont {P.~M.}\ \bibnamefont {Crespo}},\ and\ \bibinfo {author} {\bibfnamefont {J.}~\bibnamefont {{Etxezarreta Martinez}}},\ }\href@noop {} {\bibinfo {title} {Decoding algorithms for surface codes}} (\bibinfo {year} {2023}{\natexlab{a}}),\ \Eprint {https://arxiv.org/abs/2307.14989} {arXiv:2307.14989 [quant-ph]} \BibitemShut {NoStop}%
\bibitem [{\citenamefont {Shor}(1995)}]{shorQEC}%
  \BibitemOpen
  \bibfield  {author} {\bibinfo {author} {\bibfnamefont {P.~W.}\ \bibnamefont {Shor}},\ }\bibfield  {title} {\bibinfo {title} {Scheme for reducing decoherence in quantum computer memory},\ }\href {https://doi.org/10.1103/PhysRevA.52.R2493} {\bibfield  {journal} {\bibinfo  {journal} {Phys. Rev. A}\ }\textbf {\bibinfo {volume} {52}},\ \bibinfo {pages} {R2493} (\bibinfo {year} {1995})}\BibitemShut {NoStop}%
\bibitem [{\citenamefont {Kitaev}(2003)}]{kitaev}%
  \BibitemOpen
  \bibfield  {author} {\bibinfo {author} {\bibfnamefont {A.}~\bibnamefont {Kitaev}},\ }\bibfield  {title} {\bibinfo {title} {Fault-tolerant quantum computation by anyons},\ }\href {https://doi.org/https://doi.org/10.1016/S0003-4916(02)00018-0} {\bibfield  {journal} {\bibinfo  {journal} {Annals of Physics}\ }\textbf {\bibinfo {volume} {303}},\ \bibinfo {pages} {2} (\bibinfo {year} {2003})}\BibitemShut {NoStop}%
\bibitem [{\citenamefont {Fowler}\ \emph {et~al.}(2012{\natexlab{a}})\citenamefont {Fowler}, \citenamefont {Mariantoni}, \citenamefont {Martinis},\ and\ \citenamefont {Cleland}}]{fowlerReview}%
  \BibitemOpen
  \bibfield  {author} {\bibinfo {author} {\bibfnamefont {A.~G.}\ \bibnamefont {Fowler}}, \bibinfo {author} {\bibfnamefont {M.}~\bibnamefont {Mariantoni}}, \bibinfo {author} {\bibfnamefont {J.~M.}\ \bibnamefont {Martinis}},\ and\ \bibinfo {author} {\bibfnamefont {A.~N.}\ \bibnamefont {Cleland}},\ }\bibfield  {title} {\bibinfo {title} {Surface codes: Towards practical large-scale quantum computation},\ }\href {https://doi.org/10.1103/PhysRevA.86.032324} {\bibfield  {journal} {\bibinfo  {journal} {Phys. Rev. A}\ }\textbf {\bibinfo {volume} {86}},\ \bibinfo {pages} {032324} (\bibinfo {year} {2012}{\natexlab{a}})}\BibitemShut {NoStop}%
\bibitem [{\citenamefont {Bonilla~Ataides}\ \emph {et~al.}(2021)\citenamefont {Bonilla~Ataides}, \citenamefont {Tuckett}, \citenamefont {Bartlett}, \citenamefont {Flammia},\ and\ \citenamefont {Brown}}]{xzzx}%
  \BibitemOpen
  \bibfield  {author} {\bibinfo {author} {\bibfnamefont {J.~P.}\ \bibnamefont {Bonilla~Ataides}}, \bibinfo {author} {\bibfnamefont {D.~K.}\ \bibnamefont {Tuckett}}, \bibinfo {author} {\bibfnamefont {S.~D.}\ \bibnamefont {Bartlett}}, \bibinfo {author} {\bibfnamefont {S.~T.}\ \bibnamefont {Flammia}},\ and\ \bibinfo {author} {\bibfnamefont {B.~J.}\ \bibnamefont {Brown}},\ }\bibfield  {title} {\bibinfo {title} {The xzzx surface code},\ }\href {https://doi.org/10.1038/s41467-021-22274-1} {\bibfield  {journal} {\bibinfo  {journal} {Nature Communications}\ }\textbf {\bibinfo {volume} {12}},\ \bibinfo {pages} {2172} (\bibinfo {year} {2021})}\BibitemShut {NoStop}%
\bibitem [{\citenamefont {Stassi}\ \emph {et~al.}(2020)\citenamefont {Stassi}, \citenamefont {Cirio},\ and\ \citenamefont {Nori}}]{conncetSuper}%
  \BibitemOpen
  \bibfield  {author} {\bibinfo {author} {\bibfnamefont {R.}~\bibnamefont {Stassi}}, \bibinfo {author} {\bibfnamefont {M.}~\bibnamefont {Cirio}},\ and\ \bibinfo {author} {\bibfnamefont {F.}~\bibnamefont {Nori}},\ }\bibfield  {title} {\bibinfo {title} {Scalable quantum computer with superconducting circuits in the ultrastrong coupling regime},\ }\href {https://doi.org/10.1038/s41534-020-00294-x} {\bibfield  {journal} {\bibinfo  {journal} {npj Quantum Information}\ }\textbf {\bibinfo {volume} {6}},\ \bibinfo {pages} {67} (\bibinfo {year} {2020})}\BibitemShut {NoStop}%
\bibitem [{\citenamefont {Krinner}\ \emph {et~al.}(2022)\citenamefont {Krinner}, \citenamefont {Lacroix}, \citenamefont {Remm}, \citenamefont {Di~Paolo}, \citenamefont {Genois}, \citenamefont {Leroux}, \citenamefont {Hellings}, \citenamefont {Lazar}, \citenamefont {Swiadek}, \citenamefont {Herrmann}, \citenamefont {Norris} \emph {et~al.}}]{wallraffSurf}%
  \BibitemOpen
  \bibfield  {author} {\bibinfo {author} {\bibfnamefont {S.}~\bibnamefont {Krinner}}, \bibinfo {author} {\bibfnamefont {N.}~\bibnamefont {Lacroix}}, \bibinfo {author} {\bibfnamefont {A.}~\bibnamefont {Remm}}, \bibinfo {author} {\bibfnamefont {A.}~\bibnamefont {Di~Paolo}}, \bibinfo {author} {\bibfnamefont {E.}~\bibnamefont {Genois}}, \bibinfo {author} {\bibfnamefont {C.}~\bibnamefont {Leroux}}, \bibinfo {author} {\bibfnamefont {C.}~\bibnamefont {Hellings}}, \bibinfo {author} {\bibfnamefont {S.}~\bibnamefont {Lazar}}, \bibinfo {author} {\bibfnamefont {F.}~\bibnamefont {Swiadek}}, \bibinfo {author} {\bibfnamefont {J.}~\bibnamefont {Herrmann}}, \bibinfo {author} {\bibfnamefont {G.~J.}\ \bibnamefont {Norris}}, \emph {et~al.},\ }\bibfield  {title} {\bibinfo {title} {Realizing repeated quantum error correction in a distance-three surface code},\ }\href {https://doi.org/10.1038/s41586-022-04566-8} {\bibfield  {journal} {\bibinfo  {journal} {Nature}\ }\textbf {\bibinfo {volume} {605}},\ \bibinfo {pages} {669}
  (\bibinfo {year} {2022})}\BibitemShut {NoStop}%
\bibitem [{\citenamefont {Acharya}\ \emph {et~al.}(2023)\citenamefont {Acharya}, \citenamefont {Aleiner}, \citenamefont {Allen}, \citenamefont {Andersen}, \citenamefont {Ansmann}, \citenamefont {Arute}, \citenamefont {Arya}, \citenamefont {Asfaw}, \citenamefont {Atalaya}, \citenamefont {Babbush}, \citenamefont {Bacon} \emph {et~al.}}]{googleSurf}%
  \BibitemOpen
  \bibfield  {author} {\bibinfo {author} {\bibfnamefont {R.}~\bibnamefont {Acharya}}, \bibinfo {author} {\bibfnamefont {I.}~\bibnamefont {Aleiner}}, \bibinfo {author} {\bibfnamefont {R.}~\bibnamefont {Allen}}, \bibinfo {author} {\bibfnamefont {T.~I.}\ \bibnamefont {Andersen}}, \bibinfo {author} {\bibfnamefont {M.}~\bibnamefont {Ansmann}}, \bibinfo {author} {\bibfnamefont {F.}~\bibnamefont {Arute}}, \bibinfo {author} {\bibfnamefont {K.}~\bibnamefont {Arya}}, \bibinfo {author} {\bibfnamefont {A.}~\bibnamefont {Asfaw}}, \bibinfo {author} {\bibfnamefont {J.}~\bibnamefont {Atalaya}}, \bibinfo {author} {\bibfnamefont {R.}~\bibnamefont {Babbush}}, \bibinfo {author} {\bibfnamefont {D.}~\bibnamefont {Bacon}}, \emph {et~al.},\ }\bibfield  {title} {\bibinfo {title} {Suppressing quantum errors by scaling a surface code logical qubit},\ }\href {https://doi.org/10.1038/s41586-022-05434-1} {\bibfield  {journal} {\bibinfo  {journal} {Nature}\ }\textbf {\bibinfo {volume} {614}},\ \bibinfo {pages} {676} (\bibinfo {year}
  {2023})}\BibitemShut {NoStop}%
\bibitem [{\citenamefont {Bluvstein}\ \emph {et~al.}(2023)\citenamefont {Bluvstein}, \citenamefont {Evered}, \citenamefont {Geim}, \citenamefont {Li}, \citenamefont {Zhou}, \citenamefont {Manovitz}, \citenamefont {Ebadi}, \citenamefont {Cain}, \citenamefont {Kalinowski}, \citenamefont {Hangleiter}, \citenamefont {Ataides}, \citenamefont {Maskara}, \citenamefont {Cong}, \citenamefont {Gao}, \citenamefont {Rodriguez}, \citenamefont {Karolyshyn}, \citenamefont {Semeghini}, \citenamefont {Gullans}, \citenamefont {Greiner}, \citenamefont {Vuleti{\'{c}}},\ and\ \citenamefont {Lukin}}]{LukinQEC}%
  \BibitemOpen
  \bibfield  {author} {\bibinfo {author} {\bibfnamefont {D.}~\bibnamefont {Bluvstein}}, \bibinfo {author} {\bibfnamefont {S.~J.}\ \bibnamefont {Evered}}, \bibinfo {author} {\bibfnamefont {A.~A.}\ \bibnamefont {Geim}}, \bibinfo {author} {\bibfnamefont {S.~H.}\ \bibnamefont {Li}}, \bibinfo {author} {\bibfnamefont {H.}~\bibnamefont {Zhou}}, \bibinfo {author} {\bibfnamefont {T.}~\bibnamefont {Manovitz}}, \bibinfo {author} {\bibfnamefont {S.}~\bibnamefont {Ebadi}}, \bibinfo {author} {\bibfnamefont {M.}~\bibnamefont {Cain}}, \bibinfo {author} {\bibfnamefont {M.}~\bibnamefont {Kalinowski}}, \bibinfo {author} {\bibfnamefont {D.}~\bibnamefont {Hangleiter}}, \bibinfo {author} {\bibfnamefont {J.~P.~B.}\ \bibnamefont {Ataides}}, \bibinfo {author} {\bibfnamefont {N.}~\bibnamefont {Maskara}}, \bibinfo {author} {\bibfnamefont {I.}~\bibnamefont {Cong}}, \bibinfo {author} {\bibfnamefont {X.}~\bibnamefont {Gao}}, \bibinfo {author} {\bibfnamefont {P.~S.}\ \bibnamefont {Rodriguez}}, \bibinfo {author} {\bibfnamefont
  {T.}~\bibnamefont {Karolyshyn}}, \bibinfo {author} {\bibfnamefont {G.}~\bibnamefont {Semeghini}}, \bibinfo {author} {\bibfnamefont {M.~J.}\ \bibnamefont {Gullans}}, \bibinfo {author} {\bibfnamefont {M.}~\bibnamefont {Greiner}}, \bibinfo {author} {\bibfnamefont {V.}~\bibnamefont {Vuleti{\'{c}}}},\ and\ \bibinfo {author} {\bibfnamefont {M.~D.}\ \bibnamefont {Lukin}},\ }\bibfield  {title} {\bibinfo {title} {Logical quantum processor based on reconfigurable atom arrays},\ }\bibfield  {journal} {\bibinfo  {journal} {Nature}\ }\href {https://doi.org/10.1038/s41586-023-06927-3} {10.1038/s41586-023-06927-3} (\bibinfo {year} {2023})\BibitemShut {NoStop}%
\bibitem [{\citenamefont {Panteleev}\ and\ \citenamefont {Kalachev}(2021)}]{panteleev1}%
  \BibitemOpen
  \bibfield  {author} {\bibinfo {author} {\bibfnamefont {P.}~\bibnamefont {Panteleev}}\ and\ \bibinfo {author} {\bibfnamefont {G.}~\bibnamefont {Kalachev}},\ }\bibfield  {title} {\bibinfo {title} {Degenerate {Q}uantum {LDPC} {C}odes {W}ith {G}ood {F}inite {L}ength {P}erformance},\ }\href {https://doi.org/10.22331/q-2021-11-22-585} {\bibfield  {journal} {\bibinfo  {journal} {{Quantum}}\ }\textbf {\bibinfo {volume} {5}},\ \bibinfo {pages} {585} (\bibinfo {year} {2021})}\BibitemShut {NoStop}%
\bibitem [{\citenamefont {Panteleev}\ and\ \citenamefont {Kalachev}(2022)}]{panteleev2}%
  \BibitemOpen
  \bibfield  {author} {\bibinfo {author} {\bibfnamefont {P.}~\bibnamefont {Panteleev}}\ and\ \bibinfo {author} {\bibfnamefont {G.}~\bibnamefont {Kalachev}},\ }\bibfield  {title} {\bibinfo {title} {Quantum ldpc codes with almost linear minimum distance},\ }\href {https://doi.org/10.1109/TIT.2021.3119384} {\bibfield  {journal} {\bibinfo  {journal} {IEEE Transactions on Information Theory}\ }\textbf {\bibinfo {volume} {68}},\ \bibinfo {pages} {213} (\bibinfo {year} {2022})}\BibitemShut {NoStop}%
\bibitem [{\citenamefont {Bravyi}\ \emph {et~al.}(2023)\citenamefont {Bravyi}, \citenamefont {Cross}, \citenamefont {Gambetta}, \citenamefont {Maslov}, \citenamefont {Rall},\ and\ \citenamefont {Yoder}}]{bravyidecoder}%
  \BibitemOpen
  \bibfield  {author} {\bibinfo {author} {\bibfnamefont {S.}~\bibnamefont {Bravyi}}, \bibinfo {author} {\bibfnamefont {A.~W.}\ \bibnamefont {Cross}}, \bibinfo {author} {\bibfnamefont {J.~M.}\ \bibnamefont {Gambetta}}, \bibinfo {author} {\bibfnamefont {D.}~\bibnamefont {Maslov}}, \bibinfo {author} {\bibfnamefont {P.}~\bibnamefont {Rall}},\ and\ \bibinfo {author} {\bibfnamefont {T.~J.}\ \bibnamefont {Yoder}},\ }\href@noop {} {\bibinfo {title} {High-threshold and low-overhead fault-tolerant quantum memory}} (\bibinfo {year} {2023}),\ \Eprint {https://arxiv.org/abs/2308.07915} {arXiv:2308.07915 [quant-ph]} \BibitemShut {NoStop}%
\bibitem [{\citenamefont {{Xu}}\ \emph {et~al.}(2023)\citenamefont {{Xu}}, \citenamefont {{Bonilla Ataides}}, \citenamefont {{Pattison}}, \citenamefont {{Raveendran}}, \citenamefont {{Bluvstein}}, \citenamefont {{Wurtz}}, \citenamefont {{Vasic}}, \citenamefont {{Lukin}}, \citenamefont {{Jiang}},\ and\ \citenamefont {{Zhou}}}]{bonillaLDPC}%
  \BibitemOpen
  \bibfield  {author} {\bibinfo {author} {\bibfnamefont {Q.}~\bibnamefont {{Xu}}}, \bibinfo {author} {\bibfnamefont {J.~P.}\ \bibnamefont {{Bonilla Ataides}}}, \bibinfo {author} {\bibfnamefont {C.~A.}\ \bibnamefont {{Pattison}}}, \bibinfo {author} {\bibfnamefont {N.}~\bibnamefont {{Raveendran}}}, \bibinfo {author} {\bibfnamefont {D.}~\bibnamefont {{Bluvstein}}}, \bibinfo {author} {\bibfnamefont {J.}~\bibnamefont {{Wurtz}}}, \bibinfo {author} {\bibfnamefont {B.}~\bibnamefont {{Vasic}}}, \bibinfo {author} {\bibfnamefont {M.~D.}\ \bibnamefont {{Lukin}}}, \bibinfo {author} {\bibfnamefont {L.}~\bibnamefont {{Jiang}}},\ and\ \bibinfo {author} {\bibfnamefont {H.}~\bibnamefont {{Zhou}}},\ }\bibfield  {title} {\bibinfo {title} {{Constant-Overhead Fault-Tolerant Quantum Computation with Reconfigurable Atom Arrays}},\ }\href {https://doi.org/10.48550/arXiv.2308.08648} {\bibfield  {journal} {\bibinfo  {journal} {arXiv e-prints}\ ,\ \bibinfo {eid} {arXiv:2308.08648}} (\bibinfo {year} {2023})},\ \Eprint
  {https://arxiv.org/abs/2308.08648} {arXiv:2308.08648 [quant-ph]} \BibitemShut {NoStop}%
\bibitem [{\citenamefont {Dennis}\ \emph {et~al.}(2002)\citenamefont {Dennis}, \citenamefont {Kitaev}, \citenamefont {Landahl},\ and\ \citenamefont {Preskill}}]{dennis}%
  \BibitemOpen
  \bibfield  {author} {\bibinfo {author} {\bibfnamefont {E.}~\bibnamefont {Dennis}}, \bibinfo {author} {\bibfnamefont {A.}~\bibnamefont {Kitaev}}, \bibinfo {author} {\bibfnamefont {A.}~\bibnamefont {Landahl}},\ and\ \bibinfo {author} {\bibfnamefont {J.}~\bibnamefont {Preskill}},\ }\bibfield  {title} {\bibinfo {title} {Topological quantum memory},\ }\href {https://doi.org/10.1063/1.1499754} {\bibfield  {journal} {\bibinfo  {journal} {Journal of Mathematical Physics}\ }\textbf {\bibinfo {volume} {43}},\ \bibinfo {pages} {4452–4505} (\bibinfo {year} {2002})}\BibitemShut {NoStop}%
\bibitem [{\citenamefont {Wu}\ \emph {et~al.}(2022)\citenamefont {Wu}, \citenamefont {Liyanage},\ and\ \citenamefont {Zhong}}]{wuinterpretation}%
  \BibitemOpen
  \bibfield  {author} {\bibinfo {author} {\bibfnamefont {Y.}~\bibnamefont {Wu}}, \bibinfo {author} {\bibfnamefont {N.}~\bibnamefont {Liyanage}},\ and\ \bibinfo {author} {\bibfnamefont {L.}~\bibnamefont {Zhong}},\ }\href@noop {} {\bibinfo {title} {An interpretation of union-find decoder on weighted graphs}} (\bibinfo {year} {2022}),\ \Eprint {https://arxiv.org/abs/2211.03288} {arXiv:2211.03288 [quant-ph]} \BibitemShut {NoStop}%
\bibitem [{\citenamefont {Edmonds}(1965)}]{blossom1}%
  \BibitemOpen
  \bibfield  {author} {\bibinfo {author} {\bibfnamefont {J.}~\bibnamefont {Edmonds}},\ }\bibfield  {title} {\bibinfo {title} {Paths, trees, and flowers},\ }\href {https://doi.org/10.4153/CJM-1965-045-4} {\bibfield  {journal} {\bibinfo  {journal} {Canadian Journal of Mathematics}\ }\textbf {\bibinfo {volume} {17}},\ \bibinfo {pages} {449–467} (\bibinfo {year} {1965})}\BibitemShut {NoStop}%
\bibitem [{\citenamefont {Fowler}(2014)}]{fowlero1}%
  \BibitemOpen
  \bibfield  {author} {\bibinfo {author} {\bibfnamefont {A.~G.}\ \bibnamefont {Fowler}},\ }\href@noop {} {\bibinfo {title} {Minimum weight perfect matching of fault-tolerant topological quantum error correction in average $o(1)$ parallel time}} (\bibinfo {year} {2014}),\ \Eprint {https://arxiv.org/abs/1307.1740} {arXiv:1307.1740 [quant-ph]} \BibitemShut {NoStop}%
\bibitem [{\citenamefont {Fowler}\ \emph {et~al.}(2012{\natexlab{b}})\citenamefont {Fowler}, \citenamefont {Whiteside},\ and\ \citenamefont {Hollenberg}}]{fowler2}%
  \BibitemOpen
  \bibfield  {author} {\bibinfo {author} {\bibfnamefont {A.~G.}\ \bibnamefont {Fowler}}, \bibinfo {author} {\bibfnamefont {A.~C.}\ \bibnamefont {Whiteside}},\ and\ \bibinfo {author} {\bibfnamefont {L.~C.~L.}\ \bibnamefont {Hollenberg}},\ }\bibfield  {title} {\bibinfo {title} {Towards practical classical processing for the surface code},\ }\href {https://doi.org/10.1103/PhysRevLett.108.180501} {\bibfield  {journal} {\bibinfo  {journal} {Phys. Rev. Lett.}\ }\textbf {\bibinfo {volume} {108}},\ \bibinfo {pages} {180501} (\bibinfo {year} {2012}{\natexlab{b}})}\BibitemShut {NoStop}%
\bibitem [{\citenamefont {Delfosse}\ and\ \citenamefont {Nickerson}(2021)}]{UF}%
  \BibitemOpen
  \bibfield  {author} {\bibinfo {author} {\bibfnamefont {N.}~\bibnamefont {Delfosse}}\ and\ \bibinfo {author} {\bibfnamefont {N.~H.}\ \bibnamefont {Nickerson}},\ }\bibfield  {title} {\bibinfo {title} {Almost-linear time decoding algorithm for topological codes},\ }\href {https://doi.org/10.22331/q-2021-12-02-595} {\bibfield  {journal} {\bibinfo  {journal} {{Quantum}}\ }\textbf {\bibinfo {volume} {5}},\ \bibinfo {pages} {595} (\bibinfo {year} {2021})}\BibitemShut {NoStop}%
\bibitem [{\citenamefont {Delfosse}\ and\ \citenamefont {Z\'emor}(2020)}]{peeling}%
  \BibitemOpen
  \bibfield  {author} {\bibinfo {author} {\bibfnamefont {N.}~\bibnamefont {Delfosse}}\ and\ \bibinfo {author} {\bibfnamefont {G.}~\bibnamefont {Z\'emor}},\ }\bibfield  {title} {\bibinfo {title} {Linear-time maximum likelihood decoding of surface codes over the quantum erasure channel},\ }\href {https://doi.org/10.1103/PhysRevResearch.2.033042} {\bibfield  {journal} {\bibinfo  {journal} {Phys. Rev. Res.}\ }\textbf {\bibinfo {volume} {2}},\ \bibinfo {pages} {033042} (\bibinfo {year} {2020})}\BibitemShut {NoStop}%
\bibitem [{\citenamefont {Higgott}\ and\ \citenamefont {Gidney}(2023)}]{sparseblossom}%
  \BibitemOpen
  \bibfield  {author} {\bibinfo {author} {\bibfnamefont {O.}~\bibnamefont {Higgott}}\ and\ \bibinfo {author} {\bibfnamefont {C.}~\bibnamefont {Gidney}},\ }\href@noop {} {\bibinfo {title} {Sparse blossom: correcting a million errors per core second with minimum-weight matching}} (\bibinfo {year} {2023}),\ \Eprint {https://arxiv.org/abs/2303.15933} {arXiv:2303.15933 [quant-ph]} \BibitemShut {NoStop}%
\bibitem [{\citenamefont {Wu}\ and\ \citenamefont {Zhong}(2023)}]{fussionblossom}%
  \BibitemOpen
  \bibfield  {author} {\bibinfo {author} {\bibfnamefont {Y.}~\bibnamefont {Wu}}\ and\ \bibinfo {author} {\bibfnamefont {L.}~\bibnamefont {Zhong}},\ }\href@noop {} {\bibinfo {title} {Fusion blossom: Fast mwpm decoders for qec}} (\bibinfo {year} {2023}),\ \Eprint {https://arxiv.org/abs/2305.08307} {arXiv:2305.08307 [quant-ph]} \BibitemShut {NoStop}%
\bibitem [{\citenamefont {Terhal}(2015)}]{terhal}%
  \BibitemOpen
  \bibfield  {author} {\bibinfo {author} {\bibfnamefont {B.~M.}\ \bibnamefont {Terhal}},\ }\bibfield  {title} {\bibinfo {title} {Quantum error correction for quantum memories},\ }\href {https://doi.org/10.1103/RevModPhys.87.307} {\bibfield  {journal} {\bibinfo  {journal} {Rev. Mod. Phys.}\ }\textbf {\bibinfo {volume} {87}},\ \bibinfo {pages} {307} (\bibinfo {year} {2015})}\BibitemShut {NoStop}%
\bibitem [{Note1()}]{Note1}%
  \BibitemOpen
  \bibinfo {note} {Note that in the QEC jargon, the elements of the syndrome are also referred as checks. In this sense, a non-trivial check refers to a non-zero syndrome element.}\BibitemShut {Stop}%
\bibitem [{\citenamefont {Fuentes}\ \emph {et~al.}(2021)\citenamefont {Fuentes}, \citenamefont {{Etxezarreta Martinez}}, \citenamefont {Crespo},\ and\ \citenamefont {Garcia-Frías}}]{patDegen}%
  \BibitemOpen
  \bibfield  {author} {\bibinfo {author} {\bibfnamefont {P.}~\bibnamefont {Fuentes}}, \bibinfo {author} {\bibfnamefont {J.}~\bibnamefont {{Etxezarreta Martinez}}}, \bibinfo {author} {\bibfnamefont {P.~M.}\ \bibnamefont {Crespo}},\ and\ \bibinfo {author} {\bibfnamefont {J.}~\bibnamefont {Garcia-Frías}},\ }\bibfield  {title} {\bibinfo {title} {Degeneracy and its impact on the decoding of sparse quantum codes},\ }\href {https://doi.org/10.1109/ACCESS.2021.3089829} {\bibfield  {journal} {\bibinfo  {journal} {IEEE Access}\ }\textbf {\bibinfo {volume} {9}},\ \bibinfo {pages} {89093} (\bibinfo {year} {2021})}\BibitemShut {NoStop}%
\bibitem [{\citenamefont {Iyer}\ and\ \citenamefont {Poulin}(2015)}]{hardness}%
  \BibitemOpen
  \bibfield  {author} {\bibinfo {author} {\bibfnamefont {P.}~\bibnamefont {Iyer}}\ and\ \bibinfo {author} {\bibfnamefont {D.}~\bibnamefont {Poulin}},\ }\bibfield  {title} {\bibinfo {title} {Hardness of decoding quantum stabilizer codes},\ }\href {https://doi.org/10.1109/TIT.2015.2422294} {\bibfield  {journal} {\bibinfo  {journal} {IEEE Transactions on Information Theory}\ }\textbf {\bibinfo {volume} {61}},\ \bibinfo {pages} {5209} (\bibinfo {year} {2015})}\BibitemShut {NoStop}%
\bibitem [{Note2()}]{Note2}%
  \BibitemOpen
  \bibinfo {note} {Decoding stabilizer codes by taking into account all logical error classes is a \#P-complete problem \cite {hardness}.}\BibitemShut {Stop}%
\bibitem [{\citenamefont {Higgott}\ \emph {et~al.}(2023)\citenamefont {Higgott}, \citenamefont {Bohdanowicz}, \citenamefont {Kubica}, \citenamefont {Flammia},\ and\ \citenamefont {Campbell}}]{beliefmatching}%
  \BibitemOpen
  \bibfield  {author} {\bibinfo {author} {\bibfnamefont {O.}~\bibnamefont {Higgott}}, \bibinfo {author} {\bibfnamefont {T.~C.}\ \bibnamefont {Bohdanowicz}}, \bibinfo {author} {\bibfnamefont {A.}~\bibnamefont {Kubica}}, \bibinfo {author} {\bibfnamefont {S.~T.}\ \bibnamefont {Flammia}},\ and\ \bibinfo {author} {\bibfnamefont {E.~T.}\ \bibnamefont {Campbell}},\ }\bibfield  {title} {\bibinfo {title} {Improved decoding of circuit noise and fragile boundaries of tailored surface codes},\ }\href {https://doi.org/10.1103/PhysRevX.13.031007} {\bibfield  {journal} {\bibinfo  {journal} {Phys. Rev. X}\ }\textbf {\bibinfo {volume} {13}},\ \bibinfo {pages} {031007} (\bibinfo {year} {2023})}\BibitemShut {NoStop}%
\bibitem [{\citenamefont {Criger}\ and\ \citenamefont {Ashraf}(2018)}]{beliefmatchingcriger}%
  \BibitemOpen
  \bibfield  {author} {\bibinfo {author} {\bibfnamefont {B.}~\bibnamefont {Criger}}\ and\ \bibinfo {author} {\bibfnamefont {I.}~\bibnamefont {Ashraf}},\ }\bibfield  {title} {\bibinfo {title} {Multi-path summation for decoding 2d topological codes},\ }\href {https://doi.org/10.22331/q-2018-10-19-102} {\bibfield  {journal} {\bibinfo  {journal} {Quantum}\ }\textbf {\bibinfo {volume} {2}},\ \bibinfo {pages} {102} (\bibinfo {year} {2018})}\BibitemShut {NoStop}%
\bibitem [{Note3()}]{Note3}%
  \BibitemOpen
  \bibinfo {note} {Note that the number of possible branch instances cannot be higher than the number of error mechanisms. Considering that in order to start a branch instance some conditions must be met, then the actual worst case number of branch instances will be much smaller than $n$.}\BibitemShut {Stop}%
\bibitem [{\citenamefont {Kuo}\ and\ \citenamefont {Lai}(2022)}]{bpcompl}%
  \BibitemOpen
  \bibfield  {author} {\bibinfo {author} {\bibfnamefont {K.-Y.}\ \bibnamefont {Kuo}}\ and\ \bibinfo {author} {\bibfnamefont {C.-Y.}\ \bibnamefont {Lai}},\ }\bibfield  {title} {\bibinfo {title} {Exploiting degeneracy in belief propagation decoding of quantum codes},\ }\bibfield  {journal} {\bibinfo  {journal} {npj Quantum Information}\ }\textbf {\bibinfo {volume} {8}},\ \href {https://doi.org/10.1038/s41534-022-00623-2} {10.1038/s41534-022-00623-2} (\bibinfo {year} {2022})\BibitemShut {NoStop}%
\bibitem [{Note4()}]{Note4}%
  \BibitemOpen
  \bibinfo {note} {If the bottom image were to consider the left path after the first separation first and then the right one after the second, it would still require to process $5^3$ branches.}\BibitemShut {Stop}%
\bibitem [{\citenamefont {{Gidney}}(2021)}]{stim}%
  \BibitemOpen
  \bibfield  {author} {\bibinfo {author} {\bibfnamefont {C.}~\bibnamefont {{Gidney}}},\ }\bibfield  {title} {\bibinfo {title} {{Stim: a fast stabilizer circuit simulator}},\ }\href {https://doi.org/10.22331/q-2021-07-06-497} {\bibfield  {journal} {\bibinfo  {journal} {Quantum}\ }\textbf {\bibinfo {volume} {5}},\ \bibinfo {pages} {497} (\bibinfo {year} {2021})},\ \Eprint {https://arxiv.org/abs/2103.02202} {arXiv:2103.02202 [quant-ph]} \BibitemShut {NoStop}%
\bibitem [{\citenamefont {{deMarti iOlius}}\ \emph {et~al.}(2022)\citenamefont {{deMarti iOlius}}, \citenamefont {{Etxezarreta Martinez}}, \citenamefont {Fuentes}, \citenamefont {Crespo},\ and\ \citenamefont {Garcia-Frias}}]{ton1}%
  \BibitemOpen
  \bibfield  {author} {\bibinfo {author} {\bibfnamefont {A.}~\bibnamefont {{deMarti iOlius}}}, \bibinfo {author} {\bibfnamefont {J.}~\bibnamefont {{Etxezarreta Martinez}}}, \bibinfo {author} {\bibfnamefont {P.}~\bibnamefont {Fuentes}}, \bibinfo {author} {\bibfnamefont {P.~M.}\ \bibnamefont {Crespo}},\ and\ \bibinfo {author} {\bibfnamefont {J.}~\bibnamefont {Garcia-Frias}},\ }\bibfield  {title} {\bibinfo {title} {Performance of surface codes in realistic quantum hardware},\ }\href {https://doi.org/10.1103/PhysRevA.106.062428} {\bibfield  {journal} {\bibinfo  {journal} {Phys. Rev. A}\ }\textbf {\bibinfo {volume} {106}},\ \bibinfo {pages} {062428} (\bibinfo {year} {2022})}\BibitemShut {NoStop}%
\bibitem [{\citenamefont {{deMarti iOlius}}\ \emph {et~al.}(2023{\natexlab{b}})\citenamefont {{deMarti iOlius}}, \citenamefont {{Etxezarreta Martinez}}, \citenamefont {Fuentes},\ and\ \citenamefont {Crespo}}]{ton2}%
  \BibitemOpen
  \bibfield  {author} {\bibinfo {author} {\bibfnamefont {A.}~\bibnamefont {{deMarti iOlius}}}, \bibinfo {author} {\bibfnamefont {J.}~\bibnamefont {{Etxezarreta Martinez}}}, \bibinfo {author} {\bibfnamefont {P.}~\bibnamefont {Fuentes}},\ and\ \bibinfo {author} {\bibfnamefont {P.~M.}\ \bibnamefont {Crespo}},\ }\bibfield  {title} {\bibinfo {title} {Performance enhancement of surface codes via recursive minimum-weight perfect-match decoding},\ }\href {https://doi.org/10.1103/PhysRevA.108.022401} {\bibfield  {journal} {\bibinfo  {journal} {Phys. Rev. A}\ }\textbf {\bibinfo {volume} {108}},\ \bibinfo {pages} {022401} (\bibinfo {year} {2023}{\natexlab{b}})}\BibitemShut {NoStop}%
\bibitem [{\citenamefont {Tiurev}\ \emph {et~al.}(2023)\citenamefont {Tiurev}, \citenamefont {Derks}, \citenamefont {Roffe}, \citenamefont {Eisert},\ and\ \citenamefont {Reiner}}]{roffeinid}%
  \BibitemOpen
  \bibfield  {author} {\bibinfo {author} {\bibfnamefont {K.}~\bibnamefont {Tiurev}}, \bibinfo {author} {\bibfnamefont {P.-J. H.~S.}\ \bibnamefont {Derks}}, \bibinfo {author} {\bibfnamefont {J.}~\bibnamefont {Roffe}}, \bibinfo {author} {\bibfnamefont {J.}~\bibnamefont {Eisert}},\ and\ \bibinfo {author} {\bibfnamefont {J.-M.}\ \bibnamefont {Reiner}},\ }\bibfield  {title} {\bibinfo {title} {Correcting non-independent and non-identically distributed errors with surface codes},\ }\href {https://doi.org/10.22331/q-2023-09-26-1123} {\bibfield  {journal} {\bibinfo  {journal} {{Quantum}}\ }\textbf {\bibinfo {volume} {7}},\ \bibinfo {pages} {1123} (\bibinfo {year} {2023})}\BibitemShut {NoStop}%
\bibitem [{\citenamefont {Skoric}\ \emph {et~al.}(2023)\citenamefont {Skoric}, \citenamefont {Browne}, \citenamefont {Barnes}, \citenamefont {Gillespie},\ and\ \citenamefont {Campbell}}]{browneSliding}%
  \BibitemOpen
  \bibfield  {author} {\bibinfo {author} {\bibfnamefont {L.}~\bibnamefont {Skoric}}, \bibinfo {author} {\bibfnamefont {D.~E.}\ \bibnamefont {Browne}}, \bibinfo {author} {\bibfnamefont {K.~M.}\ \bibnamefont {Barnes}}, \bibinfo {author} {\bibfnamefont {N.~I.}\ \bibnamefont {Gillespie}},\ and\ \bibinfo {author} {\bibfnamefont {E.~T.}\ \bibnamefont {Campbell}},\ }\bibfield  {title} {\bibinfo {title} {Parallel window decoding enables scalable fault tolerant quantum computation},\ }\href {https://doi.org/10.1038/s41467-023-42482-1} {\bibfield  {journal} {\bibinfo  {journal} {Nature Communications}\ }\textbf {\bibinfo {volume} {14}},\ \bibinfo {pages} {7040} (\bibinfo {year} {2023})}\BibitemShut {NoStop}%
\bibitem [{\citenamefont {Tan}\ \emph {et~al.}(2023)\citenamefont {Tan}, \citenamefont {Zhang}, \citenamefont {Chao}, \citenamefont {Shi},\ and\ \citenamefont {Chen}}]{TanSliding}%
  \BibitemOpen
  \bibfield  {author} {\bibinfo {author} {\bibfnamefont {X.}~\bibnamefont {Tan}}, \bibinfo {author} {\bibfnamefont {F.}~\bibnamefont {Zhang}}, \bibinfo {author} {\bibfnamefont {R.}~\bibnamefont {Chao}}, \bibinfo {author} {\bibfnamefont {Y.}~\bibnamefont {Shi}},\ and\ \bibinfo {author} {\bibfnamefont {J.}~\bibnamefont {Chen}},\ }\bibfield  {title} {\bibinfo {title} {Scalable surface-code decoders with parallelization in time},\ }\href {https://doi.org/10.1103/PRXQuantum.4.040344} {\bibfield  {journal} {\bibinfo  {journal} {PRX Quantum}\ }\textbf {\bibinfo {volume} {4}},\ \bibinfo {pages} {040344} (\bibinfo {year} {2023})}\BibitemShut {NoStop}%
\bibitem [{\citenamefont {Chamberland}\ \emph {et~al.}(2020)\citenamefont {Chamberland}, \citenamefont {Zhu}, \citenamefont {Yoder}, \citenamefont {Hertzberg},\ and\ \citenamefont {Cross}}]{chamCLN}%
  \BibitemOpen
  \bibfield  {author} {\bibinfo {author} {\bibfnamefont {C.}~\bibnamefont {Chamberland}}, \bibinfo {author} {\bibfnamefont {G.}~\bibnamefont {Zhu}}, \bibinfo {author} {\bibfnamefont {T.~J.}\ \bibnamefont {Yoder}}, \bibinfo {author} {\bibfnamefont {J.~B.}\ \bibnamefont {Hertzberg}},\ and\ \bibinfo {author} {\bibfnamefont {A.~W.}\ \bibnamefont {Cross}},\ }\bibfield  {title} {\bibinfo {title} {Topological and subsystem codes on low-degree graphs with flag qubits},\ }\href {https://doi.org/10.1103/PhysRevX.10.011022} {\bibfield  {journal} {\bibinfo  {journal} {Phys. Rev. X}\ }\textbf {\bibinfo {volume} {10}},\ \bibinfo {pages} {011022} (\bibinfo {year} {2020})}\BibitemShut {NoStop}%
\bibitem [{\citenamefont {Chamberland}\ and\ \citenamefont {Campbell}(2022)}]{campbellCLN}%
  \BibitemOpen
  \bibfield  {author} {\bibinfo {author} {\bibfnamefont {C.}~\bibnamefont {Chamberland}}\ and\ \bibinfo {author} {\bibfnamefont {E.~T.}\ \bibnamefont {Campbell}},\ }\bibfield  {title} {\bibinfo {title} {Universal quantum computing with twist-free and temporally encoded lattice surgery},\ }\href {https://doi.org/10.1103/PRXQuantum.3.010331} {\bibfield  {journal} {\bibinfo  {journal} {PRX Quantum}\ }\textbf {\bibinfo {volume} {3}},\ \bibinfo {pages} {010331} (\bibinfo {year} {2022})}\BibitemShut {NoStop}%
\bibitem [{\citenamefont {Fuentes}\ \emph {et~al.}(2022)\citenamefont {Fuentes}, \citenamefont {{Etxezarreta Martinez}}, \citenamefont {Crespo},\ and\ \citenamefont {Garcia-Frías}}]{logicalsparse}%
  \BibitemOpen
  \bibfield  {author} {\bibinfo {author} {\bibfnamefont {P.}~\bibnamefont {Fuentes}}, \bibinfo {author} {\bibfnamefont {J.}~\bibnamefont {{Etxezarreta Martinez}}}, \bibinfo {author} {\bibfnamefont {P.~M.}\ \bibnamefont {Crespo}},\ and\ \bibinfo {author} {\bibfnamefont {J.}~\bibnamefont {Garcia-Frías}},\ }\bibfield  {title} {\bibinfo {title} {On the logical error rate of sparse quantum codes},\ }\href {https://doi.org/10.1109/TQE.2022.3196609} {\bibfield  {journal} {\bibinfo  {journal} {IEEE Transactions on Quantum Engineering}\ }\textbf {\bibinfo {volume} {3}},\ \bibinfo {pages} {1} (\bibinfo {year} {2022})}\BibitemShut {NoStop}%
\bibitem [{\citenamefont {Roffe}\ \emph {et~al.}(2020)\citenamefont {Roffe}, \citenamefont {White}, \citenamefont {Burton},\ and\ \citenamefont {Campbell}}]{roffepaperdecoding}%
  \BibitemOpen
  \bibfield  {author} {\bibinfo {author} {\bibfnamefont {J.}~\bibnamefont {Roffe}}, \bibinfo {author} {\bibfnamefont {D.~R.}\ \bibnamefont {White}}, \bibinfo {author} {\bibfnamefont {S.}~\bibnamefont {Burton}},\ and\ \bibinfo {author} {\bibfnamefont {E.}~\bibnamefont {Campbell}},\ }\bibfield  {title} {\bibinfo {title} {Decoding across the quantum low-density parity-check code landscape},\ }\href {https://doi.org/10.1103/PhysRevResearch.2.043423} {\bibfield  {journal} {\bibinfo  {journal} {Phys. Rev. Res.}\ }\textbf {\bibinfo {volume} {2}},\ \bibinfo {pages} {043423} (\bibinfo {year} {2020})}\BibitemShut {NoStop}%
\bibitem [{\citenamefont {Roffe}(2022)}]{roffbposd}%
  \BibitemOpen
  \bibfield  {author} {\bibinfo {author} {\bibfnamefont {J.}~\bibnamefont {Roffe}},\ }\href {https://github.com/quantumgizmos/bp_osd} {\bibinfo {title} {Bp+osd: A decoder for quantum ldpc codes}} (\bibinfo {year} {2022})\BibitemShut {NoStop}%
\bibitem [{\citenamefont {Jeruchim}(1984)}]{montecarlo}%
  \BibitemOpen
  \bibfield  {author} {\bibinfo {author} {\bibfnamefont {M.}~\bibnamefont {Jeruchim}},\ }\bibfield  {title} {\bibinfo {title} {Techniques for estimating the bit error rate in the simulation of digital communication systems},\ }\href {https://doi.org/10.1109/JSAC.1984.1146031} {\bibfield  {journal} {\bibinfo  {journal} {IEEE Journal on Selected Areas in Communications}\ }\textbf {\bibinfo {volume} {2}},\ \bibinfo {pages} {153} (\bibinfo {year} {1984})}\BibitemShut {NoStop}%
\end{thebibliography}%

\section*{Appendices}
\appendix

\section{BB Codes}\label{BBcodes}

In this appendix section we will describe the generation of BB codes parity check matrices through the integer values $l$ and $m$ and the polynomials $A$ and $B$. If the reader is interested in a thorough description on this class of codes the authors greatly encourage to read the seminal article on BB Codes by the IBM quantum team \cite{bravyidecoder}, this section will be heavily based on their work. 

Let us consider two integer values $l$ and $m$. We can construct two cyclic matrices $S_l$ and $S_m$, which are identity matrices of size $l\times l$ and $m \times m$ respectively with a shifting all the columns to the right one time. Given these cyclic matrices we can compute the following tensor products:

\begin{align}
\begin{split}
    x = S_l \otimes I_m
\end{split} \\
\begin{split}
    y = I_l \otimes S_m,  
\end{split},
\label{eq12}
\end{align}

where $I_k$ indicate the identity matrix of dimension $k$. In addition to the values $l$ and $m$, a BB code is also defined by a pair of matrices $A$ and $B$ which are given by two polynomials dependent on the variables $x$ and $y$ which have three non-trivial coefficients:

\begin{align}
\begin{split}
    A = A_1 + A_2 + A_3,
\end{split} \\
\begin{split}
    B = B_1 + B_2, + B_3,
\end{split}
\end{align}

where both type of matrices $A_i$ and $B_i$ take the form of 6 distinct powers of $x$ and $y$, considering that $x^l = y^m = I_{lm}$, thus, the number of possible polynomials is finite. It can be observed that the summation of the three matrices makes that for both $A$ and $B$ there are three non-trivial elements for every row and for every column. Now the $Z$ check and $X$ check parity check matrices can be constructed in the following manner:

\begin{align}
\begin{split}
    H^X = [A | B ],
\end{split} \\
\begin{split}
    H^Z = [B^T | A^T].
\end{split}
\end{align}

Due to the structure of $x$ and $y$ from eq. \eqref{eq12} we can note that they commute, thus so do $H^X$ and $H^Z$. An overall parity check matrix follows the CSS logic that $X$ errors will only be detected by the $H^Z$ submatrix and $Z$ errors will only be detected by the $H^X$ one.

\section{Noise models}\label{appNoise}
As stated in the main text, we will consider the three standard noise model abstraction levels:

In the pure data qubit noise model, the data qubits of the code are noisy while the check qubits and the stabilizer measurement circuits (as well as syndrome bits) are assumed to be noiseless \cite{decoders}. In this sense, we consider the noise model to be the standard depolarizing channel acting independently on each of the data qubits of the code. For the depolarizing channel, the probabilities of suffering a Pauli error are equiprobable, i.e. $p_x = p_y = p_z = p/3$ \cite{tvqc}.

The phenomenological noise model considers noisy data qubits as well as faulty measurement operations. In this way, the data qubits experience errors with probability $p$ in each decoding round, while the measured syndrome elements do also suffer a flip with some probability $q$. For the depolarizing phenomenological noise model, both probabilities are taken to be the same $p=q$. Therefore, this model is a second level of abstraction between considering perfect syndrome extraction circuits and taking into account all the possible error mechanisms present. Similar to the pure data qubit noise model, the occurrence of Pauli errors is taken to be equiprobable.

More generally, stabilizer measurement circuits are noisy in the reality due to faulty quantum gates and SPAM (state preparation and measurement) errors \cite{decoders}, implying that a decoder should handle such circuit-level noise for being a suitable candidate to be implemented in real quantum hardware. Here, we consider the standard depolarizing (unbiased) circuit-level noise model \cite{decoders,bravyidecoder,beliefmatching,chamCLN,campbellCLN} that consists of:
\begin{itemize}
    \item \textbf{Noisy two-qubit gates:} those are followed by a two-qubit Pauli operator, $\{\mathrm{I,X,Y,Z}\}^{\otimes 2}$, sampled independently with probability $p/15$ for the non-trivial operators and $p_\mathrm{I^{\otimes 2}} = 1-p$.
    \item \textbf{State preparation:} state preparations are followed by a Pauli operator which may flipped the state to its orthogonal one with probability $p$. Note that this reduces to substituting the preparation of the $\ket{0}$ state by $\ket{1}$ and the preparation of the $\ket{+}$ state by $\ket{-}$, each with probability $p$.
    \item \textbf{Measurements:} the outcomes of measurements are flipped with probability $p$.
    \item \textbf{Idle gate (memory) locations:} those are followed by a Pauli operator, $\{\mathrm{I,X,Y,Z}\}$, sampled independently with probabilities $p_\mathrm{X}=p_\mathrm{Y}=p_\mathrm{Z} = p/3$ and $p_\mathrm{I} = 1-p$.
\end{itemize}

\section{Numerical simulations}\label{appA}

Monte Carlo computer simulations of the BB codes have been performed with the objective of obtaining the performance curves (logical error rate) and thresholds of the code when decoded with the BPOSD and the closed-branch decoders.

The pure data qubit noise simulations in section \ref{datanoise} have been conducted in the following way. Each round of the numerical simulation is performed by generating an $N$-qubit Pauli operator, calculating its associated syndrome, and finally running the decoding algorithm using the associated syndrome as its input. Once the error is estimated by the decoder, it is used to determine if a logical error has occurred on the codestate by using the channel error. Such check is done by using the method described in \cite{logicalsparse}. 

The phenomenological and circuit-level noise simulations done in section \ref{phenomnoise} and \ref{clnnoise} have been done the following way. The sampling of the errors arising due to the noisy stabilizer circuit noise has been done by means of Stim \cite{stim}. Stim considers the check measurements upon a set of syndrome extractions altogether with a final measurement of the data qubits. We also use Stim and beliefmatching \cite{beliefmatching} for obtaining the phenomenological and circuit-level noise parity check matrices. The decoder uses those to resolve the syndrome and return an error, which is later compared to the Stim error.

The operational figure of merit we use to evaluate the performance of these quantum error correction schemes is the Logical Error Rate per syndrome cycle ($P_L$), i.e. the probability that a logical error has occurred after the recovery operation per syndrome extraction round \cite{bravyidecoder}.

Regarding the software implementations of the decoders used perform the numerical simulations have been: the BP+OSD implementation by Joschka Roffe for the BPOSD decoder \cite{roffepaperdecoding,roffbposd} (with slight modifications for handling circuit-level noise \cite{bravyidecoder}) and our implementation for the proposed closed-tree decoder will be made open source in future versions of this work.

For the numerical Monte Carlo methods employed to estimate the logical error probability per syndrome cycle, $P_L$, we have applied the following rule of thumb to select the number of simulation runs, $N_{\mathrm{runs}}$ \cite{montecarlo}, to get the logical error rate for $d$ syndrome extraction rounds, $P_L(d)$, as

\begin{equation}
N_{\mathrm{runs}} = \frac{100}{P_L(d)}.
\end{equation}

As explained in \cite{montecarlo}, under the assumption that the observed error events are independent, this results in a $95\%$ confidence interval of about $(0.8\hat{P_L(d)} , 1.25\hat{P_L(d)})$, where $\hat{P_L(d)}$ refers to the empirically estimated value for the logical error rate. The final $P_L$ is just obtained by diving the numerically estimated value by the number of syndrome extraction cycles, i.e. $P_L=P_L(d)/d$ \cite{bravyidecoder}.

\end{document}